\documentclass[journal]{IEEEtran}
\usepackage{hyperref}
\hypersetup{
	colorlinks=true,
	linkcolor=black,     
	urlcolor=black,
	citecolor = black
}
\usepackage{bbm}
\usepackage{caption}
\usepackage{amsthm}
\usepackage{utfsym}
\usepackage{amssymb,graphicx,lineno}
\usepackage{rawfonts,amsfonts,amssymb,psfrag,color}
\usepackage{amsfonts}
\usepackage{amsmath,epsfig}
\usepackage{rawfonts,graphicx,amsfonts,amssymb}
\usepackage{bm}
\usepackage{algorithm, algorithmic}
\usepackage{subfig}
\usepackage{graphicx}
\usepackage{float}
\usepackage{verbatim}
\usepackage{color,xcolor}

\newtheorem{mypro}{Proposition}

\newtheorem{lemma}{Lemma}
\newtheorem{myDef}{Definition}
\newtheorem{theorem}{Theorem}

\modulolinenumbers[5]
\usepackage{multirow}
\usepackage{booktabs}
\usepackage{cite}
\usepackage{color}

\begin{document}
	
	\title{Robust Low-Rank Matrix Completion via a New Sparsity-Inducing Regularizer}
	\author{Zhi-Yong Wang, Hing Cheung So,~\IEEEmembership{Fellow,~IEEE} and Abdelhak M. Zoubir,~\IEEEmembership{Fellow,~IEEE}
		
		\thanks{Z.-Y. Wang and H. C. So are with the Department of Electrical Engineering, City University of Hong Kong, Hong Kong, China. A. M. Zoubir is with the Signal Processing Group at Technische Universit\"at Darmstadt, 64283 Darmstadt, Germany.
			(E-mail: z.y.wang@my.cityu.edu.hk, hcso@ee.cityu.edu.hk, zoubir@spg.tu-darmstadt.de). }}
	
	\maketitle
	
	\begin{abstract}
		This paper presents a novel loss function referred to as hybrid ordinary-Welsch (HOW) and a new sparsity-inducing regularizer associated with HOW. We theoretically show that the regularizer is quasiconvex and that the corresponding Moreau envelope is convex. Moreover, the closed-form solution to its Moreau envelope, namely, the proximity operator, is derived. Compared with nonconvex regularizers like the $\ell_p$-norm with $0<p<1$ that requires iterations to find the corresponding proximity operator, the developed regularizer has a closed-form proximity operator. We apply our regularizer to the robust matrix completion problem, and develop an efficient algorithm based on the alternating direction method of multipliers. The convergence of the suggested method is analyzed and we prove that any generated accumulation point is a stationary point. Finally, experimental results based on synthetic and real-world datasets demonstrate that our algorithm is superior to the state-of-the-art methods in terms of restoration performance. 
	\end{abstract}
	
	\begin{IEEEkeywords}
		Low-rank, concave, sparsity, proximity operator, robust matrix completion.
		
	\end{IEEEkeywords}

	\section{Introduction}
	\IEEEPARstart{L}{ow-rank} matrix completion (LRMC) aims to find the missing entries of an incomplete matrix using the low-rank property~\cite{NieTC2022,ChiYN2019,MumaM2019}. The observed data in many real-life applications such as image inpainting~\cite{HuY2013,ZengWJO2018}, hyperspectral image restoration~\cite{LiXPTC2022,WangZY2022} and collaborative filtering~\cite{ZhaoLE2016,Ramlatchan2018}, may be incomplete. Thus LRMC is widely used as an efficient tool to deal with the above issues because their main information lies in a low-dimensional subspace~\cite{Davenport2016}. 
	
	Roughly speaking, LRMC can be achieved in two ways, namely, matrix factorization~\cite{ZhouTT2013,ShenY2014} and rank minimization~\cite{CandesEJ2011,Cand2009}. The former exploits LRMC via considering the estimated matrix as a product of two much smaller matrices. Much success has been reported in collaborative filtering and hyperspectral imaging with the development of efficient algorithms, including low-rank matrix fitting~\cite{WenZ2012} and alternating minimization for matrix completion~\cite{Jain2013}. Furthermore, to resist outliers, techniques such as robust matrix factorization by majorization minimization~\cite{LinZhou-2018}, practical low-rank matrix approximation via robust $\ell_1$-norm ($\rm RegL_1$) ~\cite{ZhengYG2012} and half-quadratic alternating steepest descent (HQ-ASD)~\cite{HeY2019} are proposed. Nevertheless, this approach requires knowledge of the matrix rank, which is not easy to be determined in real-world applications.
	
	Unlike matrix factorization, the rank minimization approach does not need the rank of the observed matrix. The corresponding algorithms perform LRMC via imposing a rank constraint on the estimated matrix. Because direct rank minimization is an NP-hard problem~\cite{CandesEJ2011,Cand2009}, nuclear norm minimization (NNM) as the tightest convex relaxation of rank minimization is exploited in~\cite{CandesEJ2011}. Other techniques such as singular value thresholding (SVT)~\cite{Cai2010} and accelerated proximal gradient with linesearch~\cite{Toh2010} are developed. However, NNM based algorithms shrink all singular values equally and underestimate the larger singular values~\cite{WenF2020,NieF2019}. There are two schemes to cope with such an issue. The first one is to weigh the singular values per iteration differently, which is analogous to reweighting the $\ell_1$-norm for compressive sensing~\cite{CandesEen2008}. For example, Gu $et$ $al$.~\cite{GuSH2017} propose a weighted nuclear norm minimization (WNNM) for matrix completion. They obtain good experimental results in image inpainting, although their approach is not robust against outliers. Besides, Pokala $et$ $al$.~\cite{PokalaPK2022} unfold the minimax-concave penalty (MCP)~\cite{ZhangC2010} and develop a weighted MCP (WMCP) to find the low-rank matrix. 
	
	On the other hand, nonconvex sparsity-inducing regularizers have been suggested since they have less estimation bias than the $\ell_1$-norm~\cite{WenF2020}. Various algorithms~\cite{NieFL2012,LiuLE2014,LuC2016,LuC2014,LuC2015,GongP2013} try to replace the nuclear norm with nonconvex relaxation, and have shown their superiority over NNM. {As a generalization of the nuclear norm, the Schatten $p$-norm defined as the $\ell_p$-norm of the singular values, is exploited to find the low-rank component in \cite{NieFL2012} and \cite{LiuLE2014}, and the estimation bias decreases with the $p$ value.} Lu $et$ $al$.~\cite{LuC2016,LuC2014,LuC2015} exploit nonconvex regularizers, including the exponential-type penalty~\cite{GaoC2011} and the Laplace function~\cite{TrzaskoJ2009} via iteratively reweighted nuclear norm. They attain low-rank matrix recovery, and propose generalized singular value thresholding (GSVT), which provides theoretical analysis of the low-rank optimization problem using nonconvex sparsity-promoting regularizers. In addition, the nonconvex logarithm penalty is applied to LRMC in~\cite{NieF2019}. However, the above methods are sensitive to gross errors, and impulsive noise occurs in many real-world scenarios~\cite{ZoubirAM2018,ZoubirAM2012}. {To achieve outlier resistance, Nie $et$ $al$.~\cite{NieFrobust2013,NieFrobust2012} employ joint Schatten $p$-norm and $\ell_p$-norm to model the rank minimization problem and combat gross errors, respectively.} Nevertheless, there are two main issues when the $\ell_p$-norm with $0<p<1$ is used: (i) It is not easy to choose a proper value of $p$ which is sensitive to the intensity of noise; (ii) The $\ell_p$-norm does not have a closed-form expression for its proximity operator, except for $p=\{\frac{1}{2},\frac{2}{3}\}$~\cite{KrishnanD2009}, that is, their algorithm needs iterations to find the solution to the proximity operator. To avoid iterations, two efficient $\ell_p$-norm based algorithms with $p=\frac{1}{2}$ and $p=\frac{2}{3}$, referred to as $(\pmb S$+$\pmb L)_{{1}/{2}}$ and $(\pmb S$+$\pmb L)_{{2}/{3}}$, respectively, are designed in~\cite{ShangFH-2018}.
	
	In fact, nonconvex loss functions such as Welsch and Cauchy, are widely utilized to achieve robust performance~\cite{HeRhalf2014,Lixuecauchy2019,HeYTC2022}, because the convex $\ell_1$-norm and Huber function are still sensitive to outliers with large magnitude. Among these nonconvex functions, Welsch function as an error measure has attained big success in robust principal component analysis (RPCA)~\cite{HeR2014}, robust matrix completion (RMC)~\cite{HeY2019} and subspace clustering~\cite{HeRsubspace2014}. Nevertheless, Welsch function has two limitations: (i) The first issue is stated by comparing the Welsch function with its Huber counterpart. Huber function attains robustness via dividing the data into two categories, namely, normal data and outlier-contaminated data. Here, normal data refer to observations without outliers but possibly contain Gaussian noise. The Huber function assigns equal weights for all normal data via the quadratic function, while assigning smaller weights to outlier-corrupted data using the $\ell_1$-norm. The advantage of the Huber function is that it only changes the weights of outlier-contaminated data, whereas the Welsch function down-weighs all observed data, including the normal data~\cite{ZoubirAM2018}; (ii) The implicit regularizer (IR) generated by Welsch function using half-quadratic optimization~\cite{HeR2014,HeRhalf2014} cannot make the solution sparse, limiting its applicability.
	
	In this paper, a novel loss function named \textbf{h}ybrid \textbf{o}rdinary-\textbf{W}elsch (HOW) is devised, where `ordinary' means the quadratic function or the $\ell_2$-norm. The new function only changes the weights of outlier-corrupted data and the IR generated by HOW is able to make the solution sparse. To the best of our knowledge, we are the first to propose a sparsity-inducing regularizer associated with the Welsch function, and a closed-form expression of its proximity operator, which avoids iterations to finding the corresponding solution. In addition, it is proved that the generated IR is quasiconvex and its $Moreau~envelope$ is convex. We apply the generated IR to the RMC problem, and develop an algorithm based on the alternating direction method of multipliers (ADMM).
	Our main contributions are summarized as follows:
	\begin{itemize}
		\item[(i)] We devise the HOW function, which alleviates the two limitations of Welsch function, whereby Welsch function is a special case of HOW.
		\item[(ii)] The IR generated by HOW can achieve sparseness, and the closed-form solution to its $Moreau~envelope$ is derived. Besides, the properties of the IR are theoretically analyzed. 
		\item[(iii)] The proposed sparsity-inducing regularizer is utilized to solve the RMC problem, and an ADMM based algorithm is suggested. All subproblems have closed-form solutions and we prove that any accumulation point is a stationary point that satisfies Karush-Kuhn-Tucker (KKT) conditions.
		\item[(iv)] Extensive experiments are conducted to compare the proposed algorithm with competing methods using synthetic and real-life data. It is demonstrated that our approach achieves better recovery performance.
	\end{itemize}
	
	The remainder of this paper is organized as follows. In Section \ref{Sec:Preliminaries}, we introduce notations and related works. The devised loss function and its IR are presented in Section \ref{Sec:new-loss-function}. In Section \ref{Sec:two-applications}, we apply HOW to RMC, and develop the ADMM based solver with convergence analysis. Numerical experiments using synthetic data as well as real-world images are provided in Section \ref{Results}. Finally, conclusions are drawn in Section \ref{Conclusion}.
	
	\section{Preliminaries}\label{Sec:Preliminaries}
	In this section, notations are provided and related works are reviewed.
	\subsection{Notations}
	Scalars, vectors and matrices are represented by italic, bold lower-case and bold upper-case letters, respectively. $\pmb A_{ij}$ stands for the $(i,j)$ entry of a matrix $\pmb A \in \mathbb R^{m\times n}$, and $(\cdot)^T$ signifies the transpose operator. We denote ${\Omega} \subset \{1,\cdots,m\}\times\{1,\cdots,n\}$ and $\Omega^c$ as the index set of the observed entries of a $m \times n$ matrix and the complement of $\Omega$, respectively. $(\cdot)_{\Omega}$ is defined as a projection operator:
	$$ \left[\pmb A_{\Omega}\right]_{ij}  = \left\{
	\begin{aligned}
		& A_{ij},\quad   {\rm if}~(i,j)\in \Omega  \\
		&0,\quad \quad   {\rm if}~(i,j)\in \Omega^c.  \\
	\end{aligned}
	\right.
	$$
	In addition, $\|\pmb A\|_F = \sqrt{\sum _{i=1}^{m}\sum_{j=1}^{n} A_{ij}^2}$ is its Frobenius norm. Given $\pmb B \in \mathbb{R}^{m\times n}$, $\left\langle \pmb A, \pmb B\right\rangle = {\rm trace}(\pmb A^T\pmb B)$ represents the Frobenius inner product of $\pmb A$ and $\pmb B$. Moreover, $ |a|$ represents the absolute value of the scalar $a$.	
	Finally, the first and second derivatives of a differentiable function $f(x)$ are denoted by $f'(x)$ and $f''(x)$, respectively, and $\partial f$ stands for the set of subgradients, which reduces to the derivative for differentiable functions.
	\subsection{Related Works}
	\subsubsection{Low-Rank Matrix Completion}
	Given the observed matrix $\pmb X_\Omega$, matrix completion can be written as a rank minimization problem:
	\begin{equation}\label{MC-1}
		\begin{split}
			&\mathop {\min}\limits_{\pmb M}~ \text{rank}(\pmb M), ~\text{s.t.} ~ \pmb M_{\Omega} = \pmb X_{\Omega}
		\end{split}    	   	
	\end{equation}
	where $\pmb M$ is the recovered/estimated matrix.
	However, (\ref{MC-1}) is an NP-hard problem. To solve it, many studies exploit nuclear norm as the tightest convex relaxation of the rank function~\cite{Cand2009}, leading to 
	\begin{equation}\label{nuclear-norm}
		\begin{split}
			&\mathop {\min}\limits_{\pmb M}~ \|{\pmb M}\|_*, ~ \text{s.t.} ~ \pmb M_{\Omega} = \pmb X_{\Omega}
		\end{split}    	   	
	\end{equation}
	where $\|\pmb M\|_*$ denotes the nuclear norm, which is the sum of singular values of $\pmb M$. Nevertheless, nuclear norm is equal to applying the $\ell_1$-norm to the singular value of a matrix, which underestimates all nonzero singular values and results in a biased solution. To alleviate such an issue, WNNM is suggested~\cite{GuSH2017}:
	\begin{equation}\label{WNNM-MC}
		\begin{split}
			&\mathop {\min}\limits_{\pmb M}~ \|{\pmb M}\|_{\pmb w,*}, ~ \text{s.t.} ~ \pmb M + \pmb S = \pmb X,~\pmb S_{\Omega}=\pmb 0
		\end{split}    	   	
	\end{equation}
	where $\|{\pmb M}\|_{\pmb w,*}=\sum_{i=1}^{r}\pmb w_i \pmb \sigma_i$ is the weighted nuclear norm, $\pmb \sigma_i$ is the $i$th singular value of $\pmb M$ and $\pmb w_i\geq0$ is a weight assigned to $\pmb \sigma_i$. However, the above algorithms are vulnerable to outliers. Then, an RMC approach based on the $\ell_p$-norm with $0<p<1$ is developed~\cite{NieFrobust2013}:
	\begin{equation}\label{LpSq_model}
		\begin{split}
			&\mathop {\min}\limits_{\pmb M}~ \|\pmb X_\Omega - \pmb M_\Omega\|_p^p + \gamma \|\pmb M\|_{S_q}^q
		\end{split}    	   	
	\end{equation}
	where $\|\pmb X_\Omega - \pmb M_\Omega\|_p^p = \sum_{i,j\in \Omega} (\pmb X_{ij}-\pmb M_{ij})^p$ and $\|\pmb M\|_{S_q}^q = \sum_{i=1}^{\min \{m,n\}} \pmb \sigma_i^q$. Nevertheless, the proximity operator for the $\ell_p$-norm does not have a closed-form expression, except for some special cases. 
	\subsubsection{Proximity Operator}
	The $Moreau~envelope$ of a regularizer $\varphi(\cdot)$ multiplied by a scalar $\lambda>0$ is defined as~\cite{CombettesP2005,BauschkeHH2011}:
	\begin{equation}\label{Def_Pro}
		\begin{split}
			\min\limits_{x}~\frac{1}{2}(x-y)^2 + \lambda\varphi(x)
		\end{split}    	   	
	\end{equation}
	whose solution is solved by the proximity operator:
	\begin{equation}\label{R-LSp}
		\begin{split}
			P_\varphi(y) := {\rm \arg}\min\limits_{x}~\frac{1}{2}(x-y)^2 + \lambda\varphi(x)
		\end{split}    	   	
	\end{equation} 
	In particular, the $Moreau~envelope$ of $|\cdot|_1$ is defined as:
	\begin{equation}\label{Moreau-L1}
		\begin{split}
			\min\limits_{y}~\frac{1}{2}(x-y)^2 + \lambda|y|_1
		\end{split}    	   	
	\end{equation}
	whose solution is:
	\begin{equation}\label{Pro-L1}
		\begin{split}
			y^* := P_{\ell_1,\lambda}(x)={\rm max}\{0,|x|-\lambda\}{\rm sign}(x)
		\end{split}    	   	
	\end{equation}
	which is called the proximity operator of $|\cdot|_1$, and also known as the soft-thresholding operator. However, the $\ell_1$-norm makes the solution have a constant bias $\lambda$, which can be calculated by the difference between the identity function and the solution. While Welsch function is suggested and its minimization is equivalent to maximizing the correntropy criterion~\cite{LiuW2007} when the Gaussian kernel is adopted, He $et$~$al.$~\cite{HeR2014} give its implicit regularizers (IRs) via half-quadratic optimization, and extend (\ref{Def_Pro}) to:
	\begin{equation}\label{Moreau-Me}
		\begin{split}
			l_{\varphi_w}(x):=\min\limits_{y}~\frac{1}{2}(x-y)^2 + \varphi_w(y)
		\end{split}    	   	
	\end{equation}
	where $l_{\varphi_w}(x)$ is the Welsch function, $\varphi_w(y)$ is the associated IR, and its expression is in general unknown. The solution for (\ref{Moreau-Me}) is:
	\begin{equation}\label{Pro-welsch}
		\begin{split}
			y^* := P_{\varphi_w}(x)= x-x\cdot e^{-x^2/\sigma^2}
		\end{split}    	   	
	\end{equation}
	
\begin{figure}[htb]
		\centering
		\includegraphics[width=7cm]{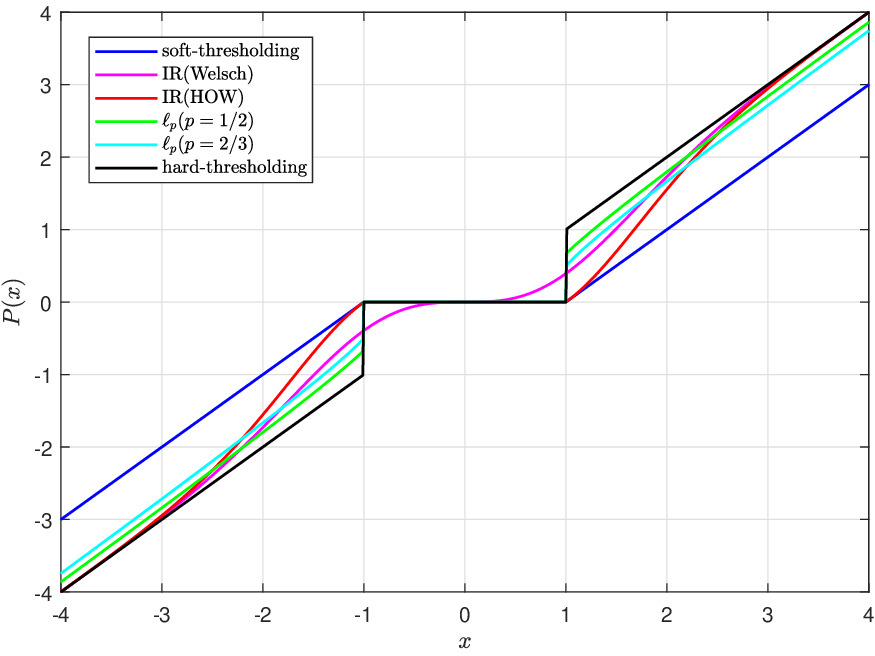}
		\vspace{-0.5em}
		\caption{Proximity operator for different regularizers with $\lambda=1$.}\label{Pro_Ope_comparison}
	\end{figure}

	Nevertheless, compared with the sparsity-promoting regularizer, such as the $\ell_p$-norm ($0<p\leq 1$), the IR of Welsch cannot produce a sparse solution for (\ref{Moreau-Me}) sparse. Fig.~\ref{Pro_Ope_comparison} shows the curves of proximity operator for different regularizers. It is observed that when $|y|\leq 1$, $P(y)=0$, that is, the regularizers $\ell_1$-norm, the IR of HOW, the $\ell_p$-norm ($0<p<1$) and the $\ell_0$-`norm' can make the solution for their corresponding optimization problem (\ref{Def_Pro}) sparse. While the solution to (\ref{Moreau-Me}) regularized by the IR of Welsch is not sparse, it is seen from (\ref{Pro-welsch}) and Fig.~\ref{Pro_Ope_comparison} that it is zero if and only if $x=0$. Moreover, the $\ell_1$-norm as a regularizer leads to a biased solution, and although the $\ell_p$-norm can alleviate this issue, the proximity operator for the $\ell_p$-norm with $0<p<1$ has no closed-form expression, except for two special cases $p=\frac{1}{2}$ and $p = \frac{2}{3}$~\cite{KrishnanD2009}, implying that iterations are needed to find its proximity operator.
	
	\section{Hybrid ordinary-Welsch function and its implicit regularizer}\label{Sec:new-loss-function}
	In this section, we first devise a novel loss function, and propose a new regularizer. We prove that the regularizer is a quasiconvex function and its $Moreau~envelope$ is convex. In addition, a closed-form expression for its proximity operator is derived.
	
	The expression of our designed HOW is:
	\begin{equation}\label{l-loss-function}
		l_{\sigma,\lambda}(x) = 
		\begin{cases}
			x^2/2, &|x|\leq \lambda\\
			\frac{\sigma^2}{2}\left(1-e^{\frac{\lambda^2-x^2}{\sigma^2}}\right)+\frac{\lambda^2}{2}, &|x|\textgreater \lambda
		\end{cases}
	\end{equation}
	where $\lambda \geq 0$ is a constant, and $\sigma$ is the kernel size. 
	It is seen that the Welsch function is a special case of (\ref{l-loss-function}) when $\lambda = 0$.
	Besides, the Legendre-Fenchel transform is utilized to study the nonconvex HOW function. Given a function $f(x)$, its conjugate $f^*(y)$ is~\cite{BoydS2004}:
	\begin{equation}\label{LF_Def}
		f^*(y) =  \mathop {\sup}\limits_{x} ~ x y - f(x)
	\end{equation}
	If $f(x)$ is a convex function, we have: 
	\begin{equation}\label{con_con_f}
		f(x)=\left(f^*(x)\right)^* =  \mathop {\max}\limits_{y} ~ xy - f^*(y)
	\end{equation}
	where the $\sup$ amounts to the $\max$ when $f(x)$ is a convex function.
	Moreover, we define a new convex function $f(x)$:
	\begin{equation}\label{fx-function}
		\begin{split}
			f(x) &=  \frac{x^2}{2} - l_{\sigma,\lambda}(x) \\
			&= 
			{\begin{cases}
					0, &|x|\leq \lambda\\
					\frac{x^2}{2} - \frac{\sigma^2}{2}\left(1-e^{\frac{\lambda^2-x^2}{\sigma^2}}\right)-\frac{\lambda^2}{2}, &|x|\textgreater \lambda
			\end{cases}}
		\end{split}
	\end{equation}
	whose convexity property is proved in Appendix~\ref{f-convex}.
	By (\ref{LF_Def}), it is easy to obtain:
	\begin{equation}\label{fy-LF}
		\begin{split}
			f^*(y) &=  \mathop {\max}\limits_{x} ~ xy - \frac{x^2}{2} + l_{\sigma,\lambda}(x)\\
			& = \mathop {\max}\limits_{x}~ -\frac{(y-x)^2}{2} + l_{\sigma,\lambda}(x) + \frac{y^2}{2}\\
			& = \lambda \varphi_{\sigma,\lambda}(y) +\frac{y^2}{2}
		\end{split}	
	\end{equation}
	where
	\begin{equation}\label{varphi-y}
		\varphi_{\sigma,\lambda}(y) = \mathop {\max}\limits_{x}~ -\frac{(y-x)^2}{2\lambda} + \frac{l_{\sigma,\lambda}(x)}{\lambda}
	\end{equation}
	Since $f(x)$ is convex, applying (\ref{con_con_f}) yields:
	\begin{equation}\label{fx-LF}
		\begin{split}
			f(x)  &=  \mathop {\max}\limits_{y} ~ y\cdot x - f^*(y) \\
			&=  \mathop {\max}\limits_{y} ~ y\cdot x - \lambda \varphi_{\sigma,\lambda}(y) - \frac{y^2}{2} \\
			& = \mathop {\max}\limits_{y}~ -\frac{(y-x)^2}{2} - \lambda \varphi_{\sigma,\lambda}(y) + \frac{x^2}{2}
		\end{split}	
	\end{equation}
	Combining (\ref{fx-function}) and (\ref{fx-LF}), we have:
	\begin{equation}\label{l-new-function}
		\begin{split}
			l_{\sigma,\lambda}(x) = \mathop {\min}\limits_{y} ~\frac{(y-x)^2}{2} + \lambda \varphi_{\sigma,\lambda}(y)
		\end{split}	
	\end{equation}
	where $\varphi_{\sigma,\lambda}(y)$ is named as the IR of HOW. Similar to the IR of the Welsch function, the exact expression of $\varphi_{\sigma,\lambda}(y)$ is unknown. 
	The solution to (\ref{l-new-function}) is the same as that to (\ref{fx-LF}), and it can be determined by the following lemma.
	\begin{lemma}\label{inversion_proximity}(Inversion rule for subgradient relations~\cite{RockafellarRT2004}) 
		For any proper, lower semicontinuous and convex function $f(x)$, we have:
		\begin{equation}\label{relationship_0}
			\begin{split}
				&\mathop {\arg \max\limits_{y}} ~ y x - f^*(y) = \partial f(x)\\
				&\mathop {\arg \max\limits_{x}} ~ x y - f(x) =  \partial f^*(y)
			\end{split}
		\end{equation}
	\end{lemma}
	Thus, the solution to (\ref{l-new-function}) is:
	\begin{equation}\label{y-pro-solution}
		P_{\varphi_{\sigma,\lambda}}(x)=f'(x) = {\rm max}\left\{0, |x|-|x|\cdot e^{(\lambda^2-x^2)/\sigma^2}\right\}{{\rm sign}(x)}
	\end{equation}
	Furthermore, the properties of $\varphi_{\sigma,\lambda}(y)$ are summarized in Proposition~\ref{solution-proximal}, whose proof is provided in Appendix~\ref{proof-solution-proximal}. 
	\begin{mypro}\label{solution-proximal} 
		$\varphi_{\sigma,\lambda}(y)$ has the following three important properties:
		\begin{itemize}
			\item[(i)] $\varphi_{\sigma,\lambda}(y)$ is concave for $y> 0$ when $\sigma \leq \sqrt{2}\lambda$, and $\varphi_{\sigma,\lambda}(y)$ is symmetric, i.e., $\varphi_{\sigma,\lambda}(y) = \varphi_{\sigma,\lambda}(-y)$. That is, $\varphi_{\sigma,\lambda}(y)$ is a quasiconvex function when $\sigma \leq \sqrt{2}\lambda$.
			\item[(ii)] Defining $g(y)=\frac{y^2}{2} + \lambda\varphi_{\sigma,\lambda}(y)$, $g(y)$ is convex with respect to (w.r.t.) $y$ for any $\lambda$ and $\sigma$.
			\item[(iii)] $P_{\varphi_{\sigma,\lambda}}(x)$ is monotonically non-decreasing, namely, for any $x_1<x_2$, $P_{\varphi_{\sigma,\lambda}}(x_1)\leq P_{\varphi_{\sigma,\lambda}}(x_2)$.
		\end{itemize}
	\end{mypro}

	It is worth pointing that although $ \varphi_{\sigma,\lambda}(y)$ is nonconvex, problem (\ref{l-new-function}) is convex due to Proposition~\ref{solution-proximal}. Fig.~\ref{Pro_Ope_comparison} plots the curve of $P_{\varphi_{\sigma,\lambda}}(x)$ with $\lambda = 1$ and $\sigma = \sqrt{2}$. It is seen that compared with the $\ell_1$-norm, the IR of HOW has a smaller bias (the bias is given by the difference between the identity function and the proximity operator for $x>\lambda$). Compared to other nonconvex regularizers such as the $\ell_p$-norm ($0<p<1$), whose corresponding optimization problems in~(\ref{Def_Pro}) may be not convex, our regularizer makes (\ref{l-new-function}) convex and its closed-form solution is derived.
	
\begin{figure}
		\centering
		\begin{minipage}{0.16\linewidth}
			\footnotesize
			\centerline{\scriptsize {$z_1 = |\pmb y_1|_1$}}
			\centerline{\includegraphics[width=3cm]{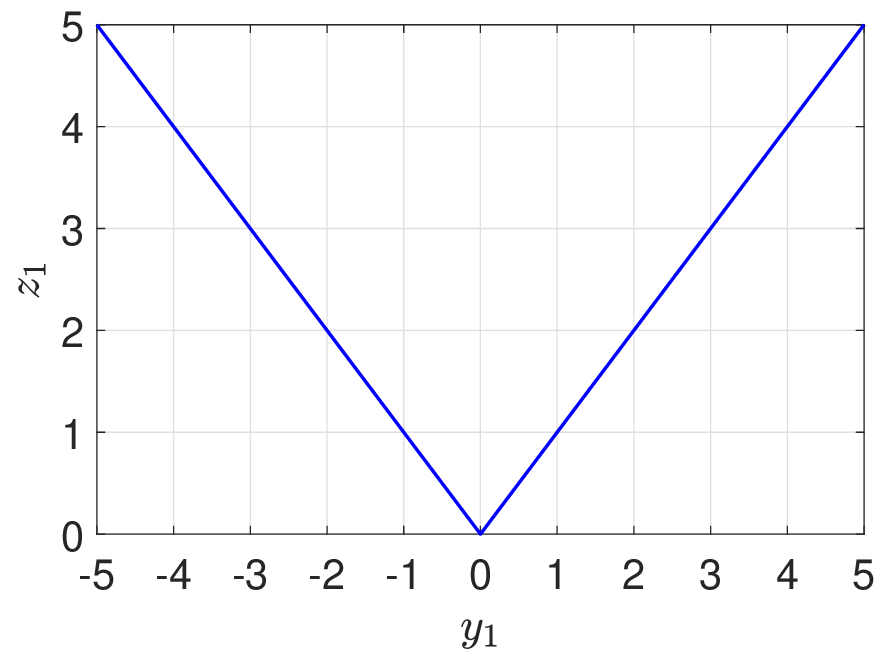}}\vskip 0pt
			\centerline{\scriptsize {(a)}}\vskip -3pt
			\centerline{ }
		\end{minipage}\hspace{15mm}
		\begin{minipage}{0.16\linewidth}
			\footnotesize
			\centerline{\scriptsize {$z_2 = \varphi_{\sigma,\lambda}(\pmb y_1)$}}
			\centerline{\includegraphics[width=3cm]{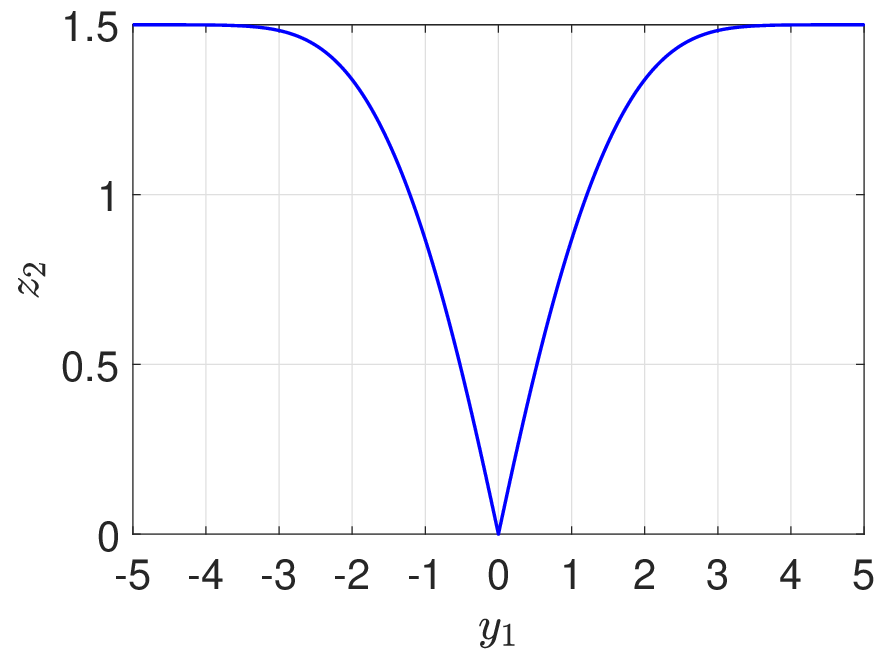}}\vskip 0pt
			\centerline{\scriptsize {(b)}}\vskip -3pt
			\centerline{ }
		\end{minipage}\hspace{15mm}
		\begin{minipage}{0.16\linewidth}
			\footnotesize
			\centerline{\scriptsize {$z_3 = g(\pmb y_1)$}}
			\centerline{\includegraphics[width=3cm]{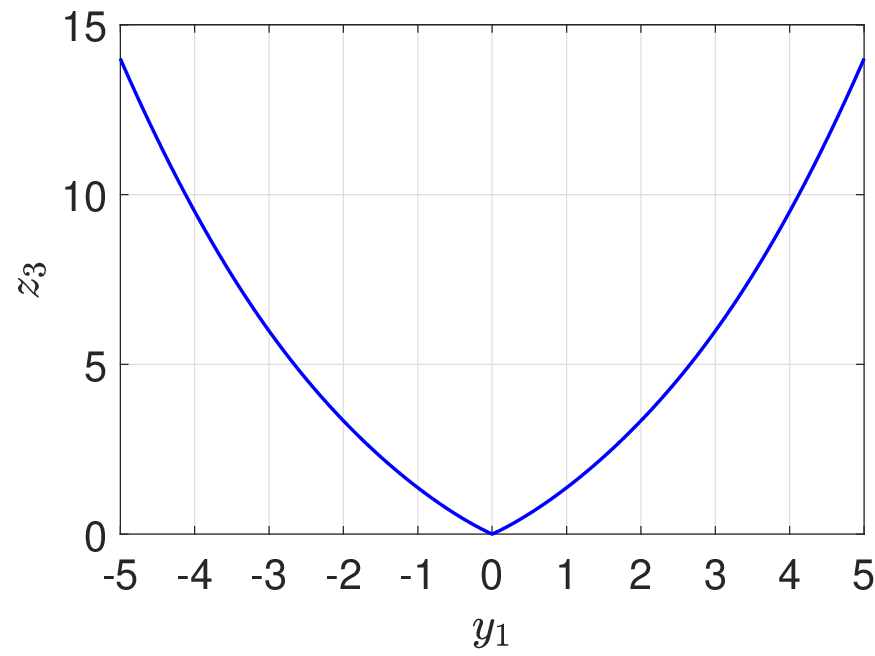}}\vskip 0pt
			\centerline{\scriptsize {(c)}}\vskip -3pt
			\centerline{ }
		\end{minipage}
		
		\begin{minipage}{0.16\linewidth}
			\footnotesize
			\centerline{\scriptsize {$z_1 = |\pmb y_1|_1+ |\pmb y_2|_1$}}
			\centerline{\includegraphics[width=2.9cm]{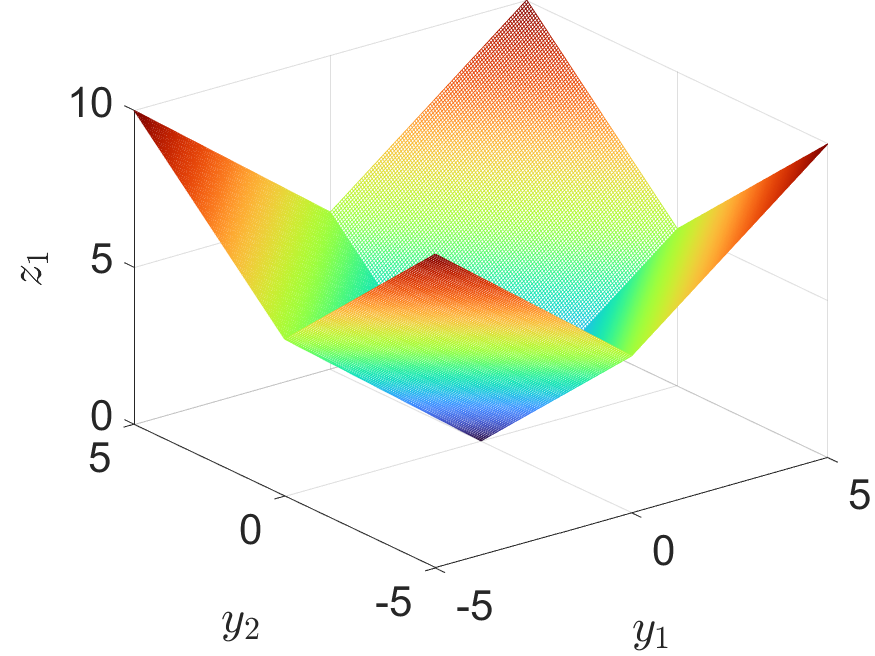}}\vskip 0pt
			\centerline{\scriptsize {(d)}}\vskip -3pt
			\centerline{ }
		\end{minipage}\hspace{15mm}
		\begin{minipage}{0.16\linewidth}
			\footnotesize
			\centerline{\scriptsize {$z_2 = \varphi_{\sigma,\lambda}(\pmb y_1)+\varphi_{\sigma,\lambda}(\pmb y_2)$}}
			\centerline{\includegraphics[width=2.9cm]{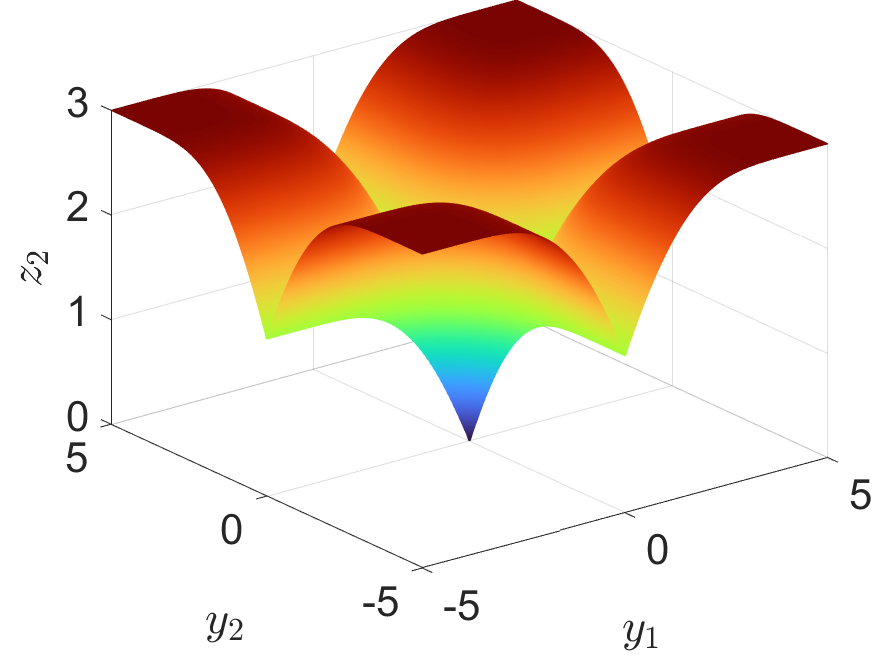}}\vskip 0pt
			\centerline{\scriptsize {(e)}}\vskip -3pt
			\centerline{ }
		\end{minipage}\hspace{15mm}
		\begin{minipage}{0.16\linewidth}
			\footnotesize
			\centerline{\scriptsize {$z_3 = g(\pmb y_1)+g(\pmb y_2)$}}
			\centerline{\includegraphics[width=2.9cm]{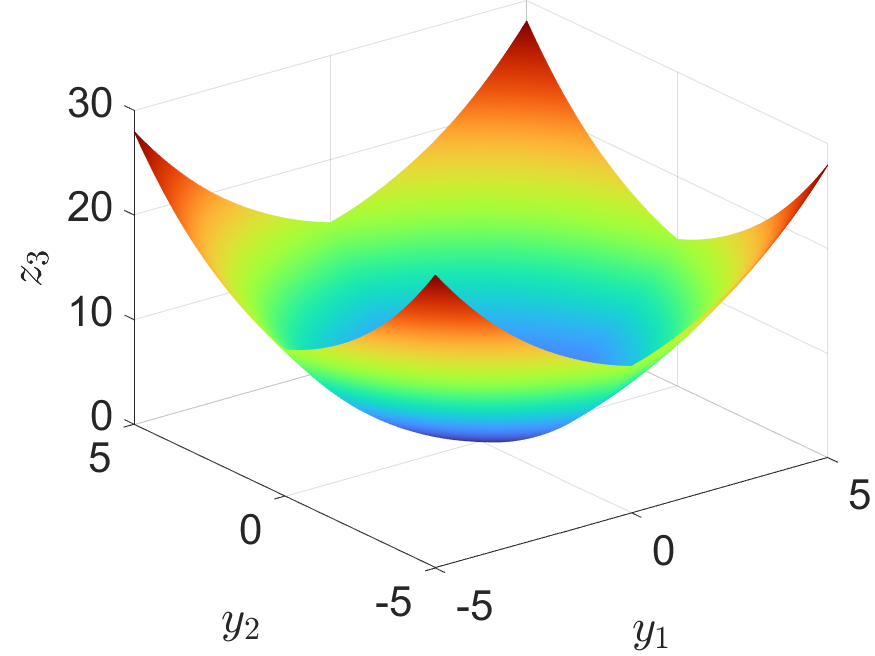}}\vskip 0pt
			\centerline{\scriptsize {(f)}}\vskip -3pt
			\centerline{ }
		\end{minipage}
		
		\begin{minipage}{0.16\linewidth}
			\footnotesize
			\centerline{\scriptsize {Contours of (d)}}
			\centerline{\includegraphics[width=2.9cm]{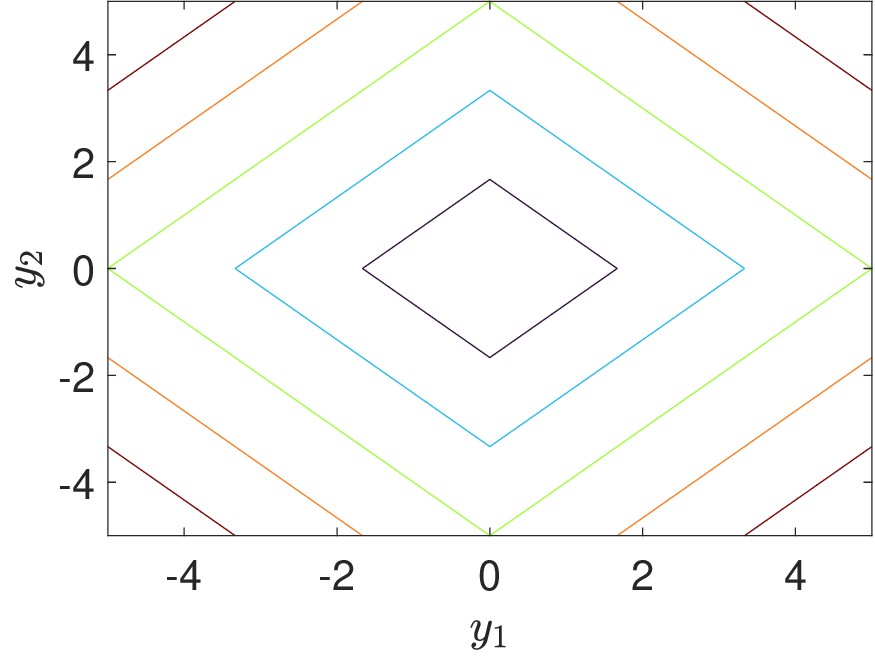}}\vskip 0pt
			\centerline{\scriptsize {(g)}}\vskip -3pt
			\centerline{ }
		\end{minipage}\hspace{15mm}
		\begin{minipage}{0.16\linewidth}
			\footnotesize
			\centerline{\scriptsize {Contours of (e)}}
			\centerline{\includegraphics[width=2.9cm]{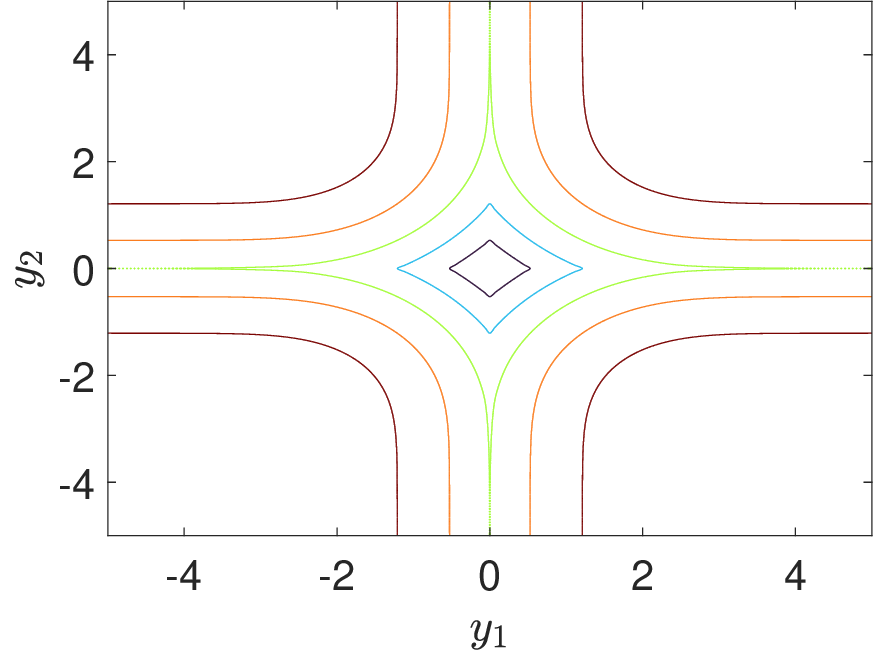}}\vskip 0pt
			\centerline{\scriptsize {(h)}}\vskip -3pt
			\centerline{ }
		\end{minipage}\hspace{15mm}
		\begin{minipage}{0.16\linewidth}
			\footnotesize
			\centerline{\scriptsize {Contours of (f)}}
			\centerline{\includegraphics[width=2.9cm]{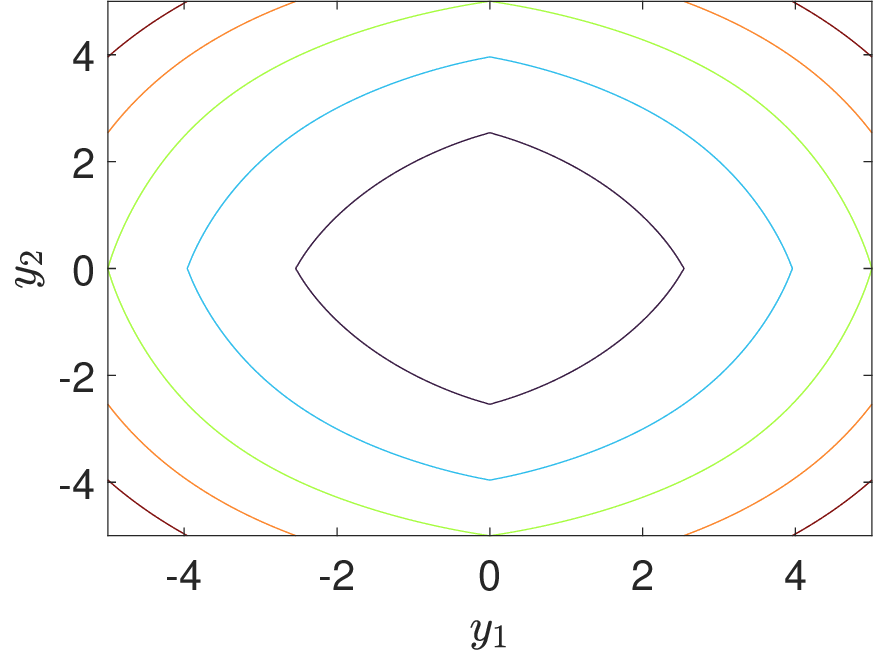}}\vskip 0pt
			\centerline{\scriptsize {(i)}}\vskip -3pt
			\centerline{ }
		\end{minipage}
		\caption{Illustration of Proposition~\ref{solution-proximal}. (a)-(c) show the curves of $|y|_1$, $\varphi_{\sigma,\lambda} (y)$ and $g( y)$, where $y$ is a scalar, which are the respective sectional views of (d)-(f). (d)-(f) plot the respective curves of $|\pmb y|_1$, $\varphi_{\sigma,\lambda} (\pmb y)$ and $g(\pmb y)$, where $\pmb y$ is a $2\times 1$ vector. (g)-(i) are the respective contours of (d)-(f).}
		\label{verify_P2}
	\end{figure}

	Moreover, the IR $\varphi_{\sigma,\lambda}(\cdot)$ is separable, that is, $\varphi_{\sigma,\lambda}(\pmb y)=\sum_{i=1}^{n} \varphi_{\sigma,\lambda}(\pmb y_i)$ where $\pmb y=[\pmb y_1,\cdots,\pmb y_n]^T$. Similarly, $g(\pmb y)=\sum_{i=1}^{i=n} g(\pmb y_i)$. To verify Proposition~\ref{solution-proximal}, Figs.~\ref{verify_P2} (d)-(f) show the curves of $|\pmb y|_1$, $\varphi_{\sigma,\lambda}(\pmb y)$ and $g(\pmb y)$ with $n=2$, respectively, with $|\pmb y|_1$ being the baseline. Figs.~\ref{verify_P2} (a)-(c) correspond to the sectional views ($\pmb y_2=0$) of (d)-(f), respectively. It is easy to see that $ \varphi_{\sigma,\lambda}(\pmb y_1)$ is concave when $\pmb y_1>0$ and $g(\pmb y_1)$ is convex. Figs.~\ref{verify_P2} (g)-(i) plot the contours of (d)-(f), respectively. We observe that the level sets (g) and (i) are convex because (d) and (f) are convex, while the level set (h) is not convex. Nevertheless, (h) can be converted into (i) via adding a quadratic term into (e). 
	
	\section{Algorithm for Robust Matrix Completion}\label{Sec:two-applications}
	\subsection{Mathematical Preliminaries}
	The key definitions and lemma used in our developed algorithm are stated in this section.
	\begin{myDef}\label{Vector_operator}
		Let $\pmb x \in \mathbb{R}^m$ and $\pmb X \in \mathbb{R}^{m\times n}$. Since the regularizer $\varphi_{\sigma,\lambda}(\cdot)$ is separable, the solution to the following problems:
		\begin{subequations}
			\begin{align}
				\mathop {\min}\limits_{\pmb y} ~\frac{1}{2}\|\pmb x -\pmb y\|_2^2 + \lambda\varphi_{\sigma,\lambda}(\pmb y)\\
				\mathop {\min}\limits_{\pmb Y} ~\frac{1}{2}\|\pmb X -\pmb Y\|_2^2 + \lambda\varphi_{\sigma,\lambda}(\pmb Y)
			\end{align}
		\end{subequations}
		are
		\begin{subequations}
			\begin{align}
				\pmb y_i = P_{\varphi_{\sigma,\lambda}}(\pmb x_i),~i=1,\cdots,m\label{y_solution}\\
				\pmb Y_{ij} = P_{\varphi_{\sigma,\lambda}}(\pmb X_{ij}),~i=1,\cdots,m,~j=1,\cdots, n\label{Y_solution}
			\end{align}
		\end{subequations}
		respectively.
		Defining $P_{\varphi_{\sigma,\lambda}}(\cdot)$ is an element-wise operator, (\ref{y_solution}) and (\ref{Y_solution}) are denoted as:
		\begin{subequations}
			\begin{align}
				\pmb y = P_{\varphi_{\sigma,\lambda}}(\pmb x) \\
				\pmb Y = P_{\varphi_{\sigma,\lambda}}(\pmb X)
			\end{align}
		\end{subequations}
	\end{myDef}
	
	\begin{myDef}\label{Nuclear_norm}
		Let $\pmb X=\pmb U~{\rm diag}(\pmb s)~\pmb V^T$ be the singular value decomposition (SVD) of a rank-$r$ matrix $\pmb X \in \mathbb{R}^{m\times n}$, where $\pmb s = [s_1, s_2,\cdots,s_r]^T$ is the vector of singular values. The nuclear norm $\|\pmb X\|_*$ is defined as:
		\begin{equation}\label{Nuclear_norm1}
			\|\pmb X\|_* = \|\pmb s\|_1=\sum_{i=1}^{r} s_i
		\end{equation}
		which is the $\ell_1$-norm of $\pmb s$.
	\end{myDef}
	
	Using the nuclear norm to find the low-rank components will underestimate all nonzero singular values because the nuclear norm is equivalent to applying the $\ell_1$-norm to the singular values. To address this issue, we replace the $\ell_1$-norm with our sparsity-promoting regularizer.
	\begin{myDef}\label{phi_norm}
		Let $\pmb X=\pmb U~{\rm diag}(\pmb s)~\pmb V^T$ be the SVD of a rank-$r$ matrix $\pmb X \in \mathbb{R}^{m\times n}$, where $\pmb s = [s_1, s_2,\cdots,s_r]^T$ is the vector of singular values. The matrix $\varphi_{\sigma,\lambda}$-norm of $\pmb X$, denoted as $\|\pmb X\|_{\varphi_{\sigma,\lambda}}$, is defined as:
		\begin{equation}\label{phi_norm1}
			\|\pmb X\|_{\varphi_{\sigma,\lambda}} = \varphi_{\sigma,\lambda}(\pmb s)=\sum_{i=1}^{r} \varphi_{\sigma,\lambda}(s_i)
		\end{equation}
	\end{myDef}
	
	\begin{lemma}~\cite{LuC2015}\label{proximal_Matrix_norm0} 
		Let $\pmb X=\pmb U~{\rm Diag}(\pmb s)~\pmb V^T$ be the SVD of a rank-$r$ matrix $\pmb X \in \mathbb{R}^{m\times n}$, where $\pmb s = [s_1, s_2,\cdots,s_r]^T$ is the vector of singular values, and define:
		\begin{equation}\label{proximal_Matrix_norm3}
			P_{\|\cdot\|_{\varphi_{\sigma,\lambda}}}(\pmb X) = {\rm arg}\mathop {\min}\limits_{\pmb M}\lambda\|\pmb M\|_{\varphi_{\sigma,\lambda}} + \frac{1}{2}\left\|\pmb X -\pmb M\right\|_F^2
		\end{equation}
		If the proximity operator $P_{\varphi_{\sigma,\lambda}}$ is monotonically non-decreasing, then the solution to (\ref{proximal_Matrix_norm3}) is:
		\begin{equation*}
			\pmb M= \pmb U {\rm Diag}(\pmb s^\star)\pmb V^T
		\end{equation*}
		where $\pmb s^\star$ satisfies $s_1^\star\geq \cdots\geq s_i^\star\geq \cdots \geq s_r^\star $, which is determined for  $i = 1,2,\cdots,r$, as:  
		\begin{equation*}
			s_i^\star:= P_{\varphi_{\sigma,\lambda}}(s_i)={\rm arg}\mathop {\min}\limits_{s>0} \lambda\varphi_{\sigma,\lambda}(s) + \frac{1}{2}\left(s -s_i\right)^2
		\end{equation*}
	\end{lemma}

	\subsection{Algorithm Development}
	In this section, we apply the proposed sparsity-inducing regularizer to RMC. The corresponding optimization problem is written as:
	\begin{equation}
		\begin{split}
			&\mathop {\min}\limits_{\pmb M, \pmb S} ~\|\pmb M\|_{\varphi_{\sigma,1/\rho}} + \lambda\varphi_{\sigma,\lambda/\rho}(\pmb S_\Omega)\\
			&~\text{s.t.}~ \pmb X_\Omega = \pmb M_\Omega + \pmb S_\Omega
		\end{split}
	\end{equation}
	which is equal to:
	\begin{equation}\label{RMC-formulation}
		\begin{split}
			&\mathop {\min}\limits_{\pmb M, \pmb S} ~\|\pmb M\|_{\varphi_{\sigma,1/\rho}} + \lambda\varphi_{\sigma,\lambda/\rho}(\pmb S_\Omega)\\
			&~\text{s.t.}~\pmb X = \pmb M + \pmb S
		\end{split}
	\end{equation}
	where $\pmb S_{ \Omega^c} \neq 0$ if $\pmb M_{ \Omega^c} \neq 0$. Problem (\ref{RMC-formulation}) can be efficiently solved by ADMM, and its augmented Lagrangian function is:
	\begin{equation}
		\begin{split}
			\mathcal{L}'_\rho(\pmb M, \pmb S,\pmb \Lambda) &:= \|\pmb M\|_{\varphi_{\sigma,1/\rho}} + \lambda\varphi_{\sigma,\lambda/\rho}(\pmb S_\Omega) \\
			&+\left<\pmb \Lambda, \pmb X-\pmb M -\pmb S\right> +\frac{\rho}{2}\left\|\pmb X-\pmb M -\pmb S\right\|_F^2
		\end{split}
	\end{equation}
	which amounts to: 
	\begin{equation}\label{Aug_L}
		\begin{split}
			\mathcal{L}_\rho(\pmb M, \pmb S,\pmb \Lambda) &:= {1}/{\rho}\cdot\|\pmb M\|_{\varphi_{\sigma,1/\rho}} + \lambda/\rho\cdot\varphi_{\sigma,\lambda/\rho}(\pmb S_\Omega) \\
			&+\left<\pmb \Lambda, \pmb X-\pmb M -\pmb S\right>/\rho +\frac{1}{2}\left\|\pmb X-\pmb M -\pmb S\right\|_F^2
		\end{split}
	\end{equation}
	where $\pmb \Lambda$ is the Lagrange multiplier vector, the last term is the augmented term and $\rho>0$ is the penalty parameter.
	
	The details of the parameter updates at the $(k+1)$th iteration, i.e., $\left(\pmb M^{k+1}, \pmb S^{k+1}, \pmb \Lambda^{k+1}\right)$, are derived as follows.
	
	$Update~of $~$\pmb M$: Given $\pmb S^{k}$, $\pmb \Lambda^k$ and $\rho^k$, the low-rank matrix $\pmb M$ is updated by:  	
	\begin{equation}\label{update_M}
		\pmb M^{k+1}= {\rm arg}\mathop {\min}\limits_{\pmb M} 1/\rho^k \cdot\|\pmb M\|_{\varphi_{\sigma,1/\rho^k}} + \frac{1}{2}\left\|\pmb X-\pmb S^k + \frac{\pmb \Lambda^k}{\rho^k} -\pmb M\right\|_F^2
	\end{equation}
	Invoking Lemma~\ref{proximal_Matrix_norm0}, we have:
	\begin{equation}\label{M_solution}
		\pmb M^{k+1}= P_{\|\cdot\|_{\varphi_{\sigma,1/\rho^k}}}\left(\pmb X-\pmb S^{k} + \frac{\pmb \Lambda^k}{\rho^k}\right)
	\end{equation}
	
	$Update~of $~$\pmb S$: Given $\pmb M^{k+1}$, $\pmb \Lambda^k$ and $\rho^k$, $\pmb S^{k+1}$ is updated by two steps, i.e., the updates of $\pmb S_{\Omega}^{k+1}$ and $\pmb S_{\Omega^c}^{k+1}$. $\pmb S_{\Omega}^{k+1}$ is obtained from:
	\begin{equation}\label{update_C_Om}
		{\rm arg}\mathop {\min}\limits_{\pmb S_\Omega} \lambda/\rho^k\cdot\varphi_{\sigma,\lambda/\rho^k}(\pmb S_\Omega) + \frac{1}{2}\left\|\pmb X_\Omega-\pmb M_\Omega^{k+1} + \frac{\pmb \Lambda_\Omega^k}{\rho^k} -\pmb S_\Omega\right\|_F^2
	\end{equation}
	whose closed-form solution is:
	\begin{equation}\label{S_Omega}
		\pmb S_\Omega^{k+1}= P_{\varphi_{\sigma,\lambda/\rho^k}}\left(\pmb X_\Omega-\pmb M^{k+1}_\Omega + \frac{\pmb \Lambda_\Omega^k}{\rho^k}\right)
	\end{equation}
	While $\pmb S_{\Omega^c}^{k+1}$ is updated by:
	\begin{equation}\label{S_COmega}
		{\rm arg}\mathop {\min}\limits_{\pmb S_{\Omega^c}} \frac{1}{2}\left\|\pmb X_{\Omega^c}-\pmb M_{\Omega^c}^{k+1} + \frac{\pmb \Lambda_{\Omega^c}^k}{\rho^k} -\pmb S_{\Omega^c}\right\|_F^2
	\end{equation}
	with the optimal solution:
	\begin{equation}\label{S_solution_Om_S}
		\pmb S_{\Omega^c}^{k+1} = \frac{\pmb \Lambda_{\Omega^c}^k}{\rho^k} -\pmb M_{\Omega^c}^{k+1}
	\end{equation} 
	Combining (\ref{S_Omega}) and (\ref{S_solution_Om_S}) yields:
	\begin{equation}\label{S_solution}
		\pmb S_{ij}^{k+1}  = \left\{
		\begin{aligned}
			& P_{\varphi_{\sigma,\lambda/\rho^k}}\left(\pmb X_{ij}-\pmb M^{k+1}_{ij} + \frac{\pmb \Lambda_{ij}^k}{\rho^k}\right),\quad   {\rm if}~(i,j)\in \Omega  \\
			& \frac{\pmb \Lambda_{ij}^k}{\rho^k} -\pmb M_{ij}^{k+1},\quad \quad           {\rm if}~(i,j)\in { \Omega^c}.  \\
		\end{aligned}
		\right.
	\end{equation}
	
	$Update~of $~$\pmb \Lambda$: Given $\pmb M^{k+1}$, $\pmb S^{k+1}$ and $\rho^k$, $\pmb \Lambda^{k+1}$ is updated according to
	\begin{equation}\label{ADMM_Ladm}
		\pmb \Lambda^{k+1}= \pmb \Lambda^{k}+\rho^k\left(\pmb X -\pmb M^{k+1}-\pmb S^{k+1}\right)
	\end{equation} 
	The penalty parameter $\rho^k$ is determined by $\rho^{k+1} = \mu\rho^k$, where $\mu>1$ is a constant.
	The steps of the proposed algorithm are summarized in Algorithm~\ref{Algo:NNSR}. 	
	\begin{algorithm}
		\caption{Robust matrix completion via nonconvex and nonsmooth sparse regularizer (NNSR)}
		\label{Algo:NNSR}
		\algsetup{indent=1.5em}
		\vspace{1ex}
		\begin{algorithmic}
			\REQUIRE  Incomplete matrix $\pmb X_\Omega$, index set $\Omega$, $\rho^0 >0$, $\mu>1$, $\xi>0$ and $I_m$
			\STATE \textbf{Initialize:} $\pmb S^0=\pmb 0$, $\pmb \Lambda^0=\pmb 0$, and $k=0$.
			\WHILE {$rel_E^k>\xi$ and $k\leq I_m$}
			
			\STATE Update $\pmb M^k$ via (\ref{M_solution})
			
			\STATE Update $\pmb S^k$ via (\ref{S_solution})	
			
			\STATE Update $\pmb \Lambda^k$ via (\ref{ADMM_Ladm})
			
			\STATE Update $\rho^{k+1}=\mu\rho^k$	
			
			\STATE $k\leftarrow k+1$
			
			\ENDWHILE
			\ENSURE $\pmb M = \pmb M^{k}$.
		\end{algorithmic}
	\end{algorithm}
	
	\subsection{Convergence Analysis}
	The convergence of the proposed algorithm is analyzed in this section and we show that any generated accumulation point satisfies the KKT conditions.
	\begin{theorem}\label{theorem-convergence} 
		Let $\{(\pmb M^k, \pmb S^k, \pmb \Lambda^k)\}$ be the sequence generated by Algorithm~\ref{Algo:NNSR}. Given a bounded initialization $(\pmb S^0, \pmb \Lambda^0)$, $\{(\pmb M^k, \pmb S^k, \pmb \Lambda^k)\}$ has the following properties:
		\begin{itemize}
			\item[(i)] The sequence $\{(\pmb M^k, \pmb S^k)\}$ satisfies:
			\begin{enumerate}
				\item
				$\lim_{k \to \infty}\left\|\pmb M^{k+1}-\pmb M^k\right\|_F^2=0$
				\item
				$\lim_{k \to \infty}\left\|\pmb S^{k+1}-\pmb S^k\right\|_F^2=0$
				\item
				$\lim_{k \to \infty}\left\|\pmb X-\pmb M^{k+1} -\pmb S^{k+1}\right\|_F^2=0$
			\end{enumerate}
			\item[(ii)]The sequences $\{(\pmb M^k, \pmb S^k, \pmb \Lambda^k)\}$ generated are all bounded.
			\item[(iii)]Any accumulation point of the iteration sequence is a stationary point that satisfies the KKT conditions for (\ref{RMC-formulation}).
		\end{itemize}
	\end{theorem}
	whose proof can be found in Appendix~\ref{proof-theorem-convergence}. 
	
	\subsection{Stopping Criteria and Computational Complexity}
	
	The algorithm is terminated when it converges or the iteration number reaches the maximum allowable number $I_m$. Defining the relative error $ rel_E^k = \|\pmb X - \pmb M^k -\pmb S^k\|_F/\|\pmb X\|_F$, if $ rel_E^k \leq \xi$, where $\xi$ is a constant, we assert that the solution satisfies the convergence condition. 
	
	Similar to principal component pursuit (PCP)~\cite{CandesEJ2011}, the proposed algorithm involves the SVD computation per iteration, whose complexity is $\mathcal{O}(\min (m,n)mn)$~\cite{ShangFH-2018}, where $m$ and $n$ are the row and column lengths of the incomplete matrix, respectively. Thus, the total complexity of Algorithm~\ref{Algo:NNSR} is $\mathcal{O}(K\min (m,n)mn)$, where $K$ is the required iteration number.

	\section{EXPERIMENTAL RESULTS}\label{Results}
	
	In this section, we evaluate the proposed algorithm on synthetic data, real-world images and multispectral images. All simulations are conducted using a computer with 3.0 GHz CPU and 16 GB memory. The algorithms based on factorization, i.e., HQ-ASD~\cite{HeY2019}, $\rm RegL_1$~\cite{ZhengYG2012}, and the rank minimization algorithms including $(\pmb S$+$\pmb L)_{{1}/{2}}$, $(\pmb S$+$\pmb L)_{{2}/{3}}$~\cite{ShangFH-2018}  and $\rm LpSq$~\cite{NieFrobust2013} with $p=1/2$, are realized as competitors.
	The recommended setting of the parameters for the competing algorithms is adopted, and we suggest $\sigma = \sqrt{2}\lambda$, $\mu=1.05$, $I_m = 1000$ and $\xi=10^{-7}$ for our method.

	\subsection{Synthetic Data}	
	We first generate the low-rank matrix $\pmb M_t=\pmb U \pmb V^T$, where the entries of $\pmb U \in \mathbb{R}^{m\times r}$ and $\pmb V \in \mathbb{R}^{n\times r}$ with $r$ being the rank are standard Gaussian distributed. Then $\pmb M_t$ is corrupted by the sparse outlier matrix $\pmb S$, which includes $\alpha mn$ nonzero outliers with values uniformly distributed in $[-\beta/2,\beta/2]$. Besides, $\pmb M_t$ is masked by $\pmb \Omega$, whose entries are drawn independently from a Bernoulli distribution with $|\pmb \Omega|_1 = \gamma mn$ where $\gamma$ is the observation ratio. The relative reconstruction error (RRE) of the low-rank matrix defined as ${\rm REE} = \|\pmb M_t - {\pmb M}\|_F^2/\left\|\pmb M_t\right\|_F^2$, where ${\pmb M}$ is the estimated low-rank matrix, is employed as the evaluation metric. Moreover, the performance of all approaches is evaluated using the average results of $100$ independent runs.
	
	\begin{figure}
		\centering
		\begin{minipage}{0.16\linewidth}
			\footnotesize
			\vspace{1pt}
			\centerline{\includegraphics[width=3cm]{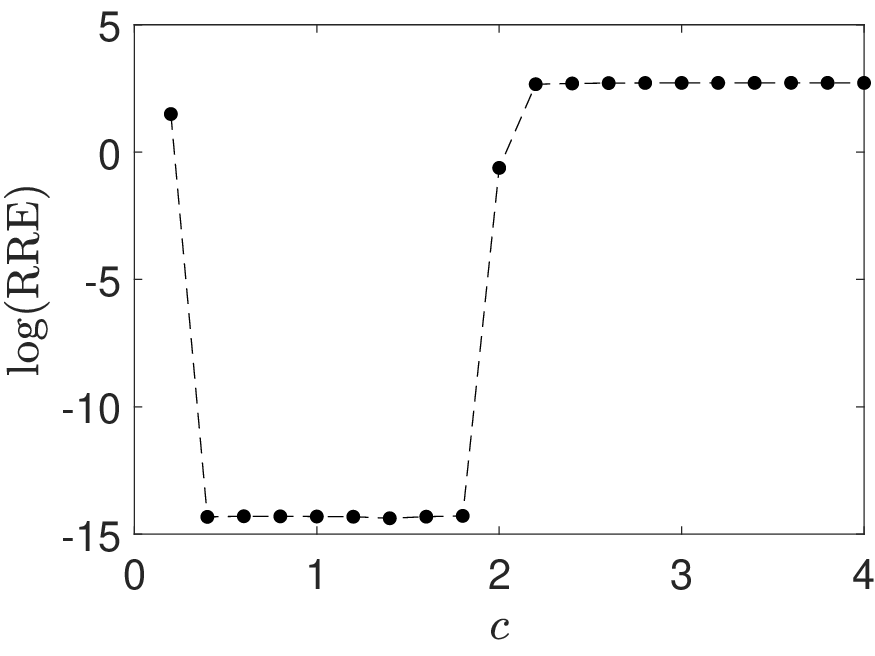}}\vskip 0pt
			\centerline{\scriptsize {(a)~$\gamma=0.9$}}\vskip -3pt
			\centerline{ }
		\end{minipage}\hspace{16mm}
		\begin{minipage}{0.16\linewidth}
			\footnotesize
			\vspace{1pt}
			\centerline{\includegraphics[width=3cm]{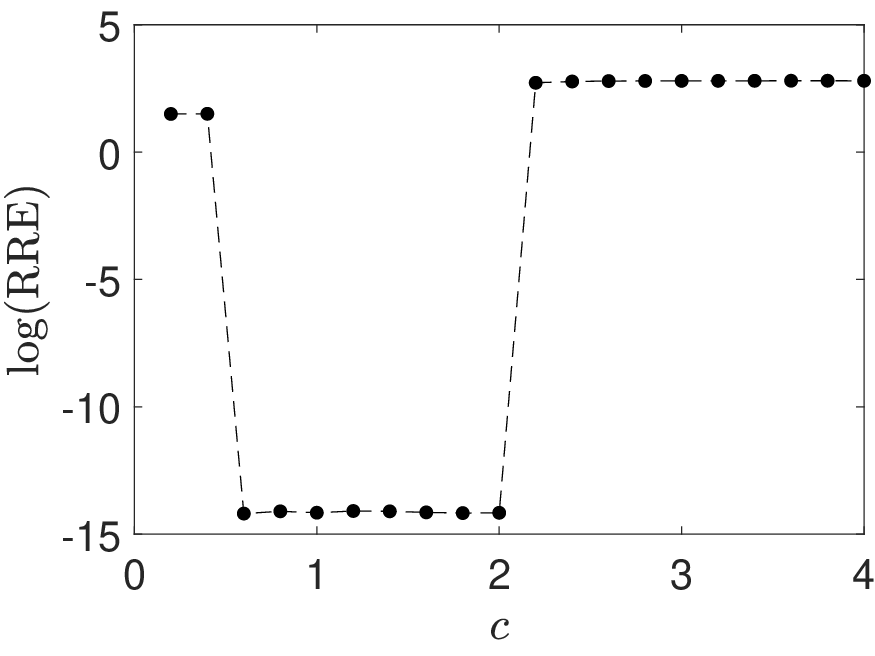}}\vskip 0pt
			\centerline{\scriptsize {(b)~$\gamma=0.7$}}\vskip -3pt
			\centerline{ }
		\end{minipage}\hspace{16mm}
		\begin{minipage}{0.16\linewidth}
			\footnotesize
			\vspace{1pt}
			\centerline{\includegraphics[width=3cm]{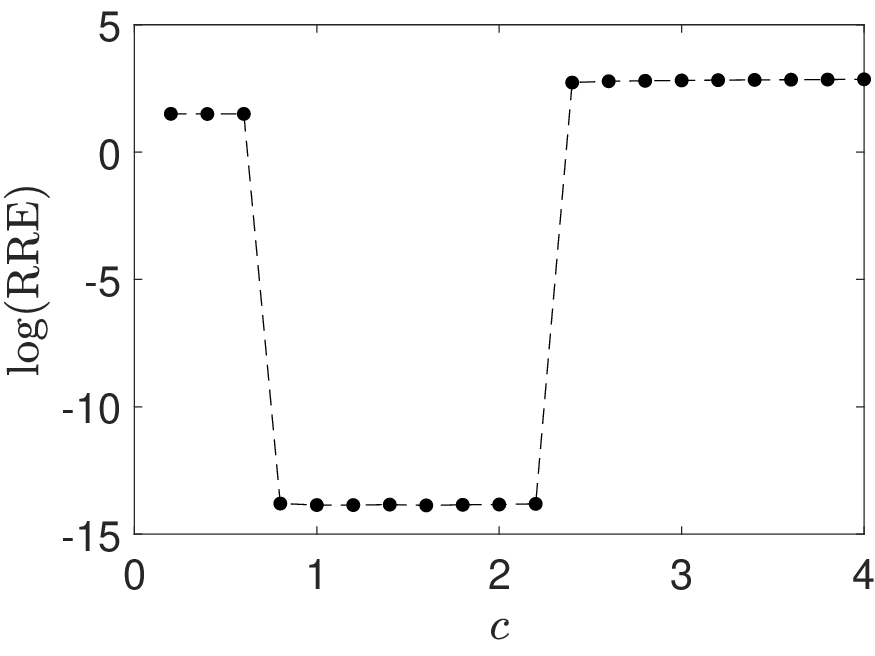}}\vskip 0pt
			\centerline{\scriptsize {(c)~$\gamma=0.5$}}\vskip -3pt
			\centerline{ }
		\end{minipage}
		
		\begin{minipage}{0.16\linewidth}
			\footnotesize
			\vspace{1pt}
			\centerline{\includegraphics[width=3cm]{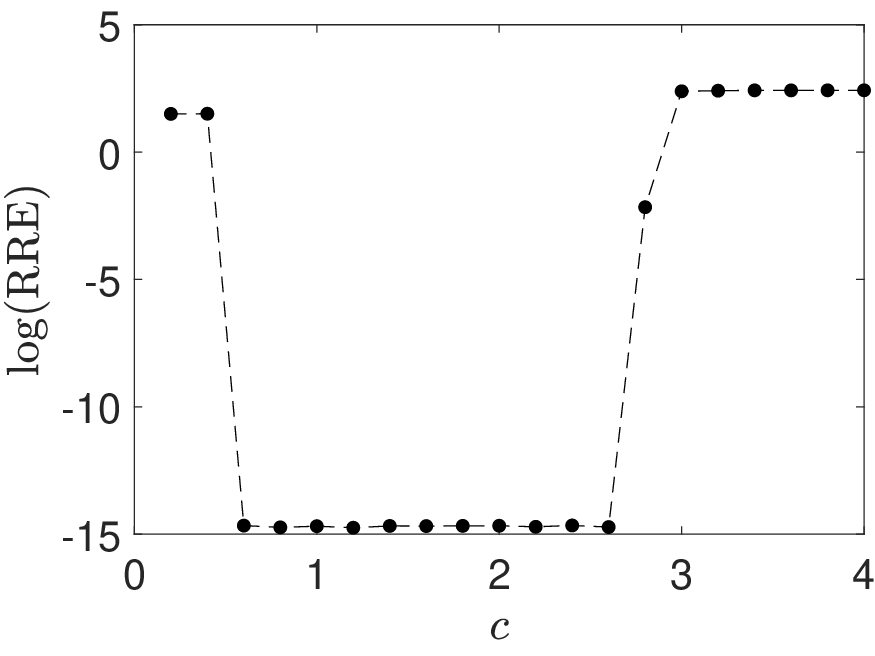}}\vskip 0pt
			\centerline{\scriptsize {(d)~$\alpha=0.1$}}\vskip -3pt
			\centerline{ }
		\end{minipage}\hspace{16mm}
		\begin{minipage}{0.16\linewidth}
			\footnotesize
			\vspace{1pt}
			\centerline{\includegraphics[width=3cm]{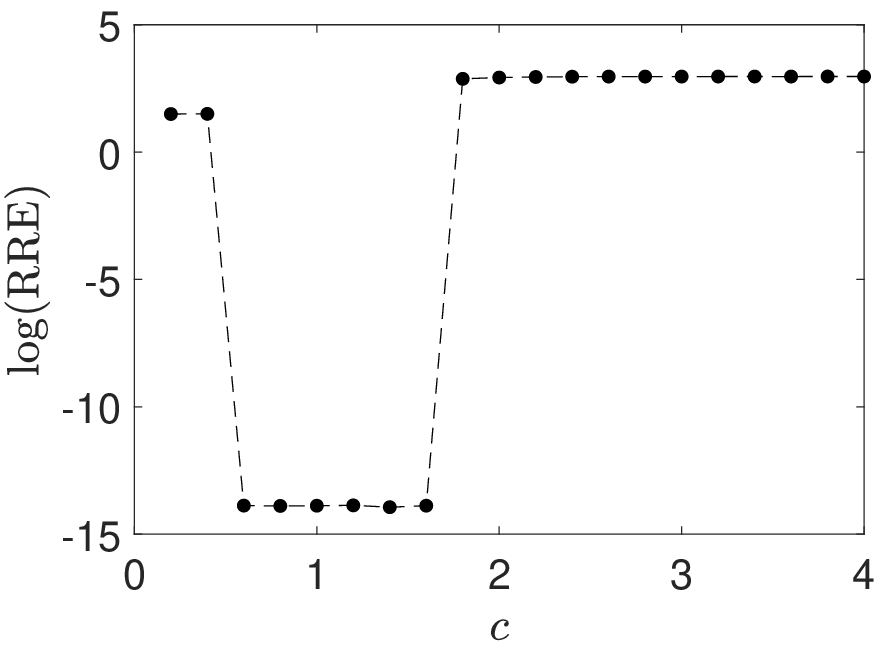}}\vskip 0pt
			\centerline{\scriptsize {(e)~$\alpha=0.3$}}\vskip -3pt
			\centerline{ }
		\end{minipage}\hspace{16mm}
		\begin{minipage}{0.16\linewidth}
			\footnotesize
			\vspace{1pt}
			\centerline{\includegraphics[width=3cm]{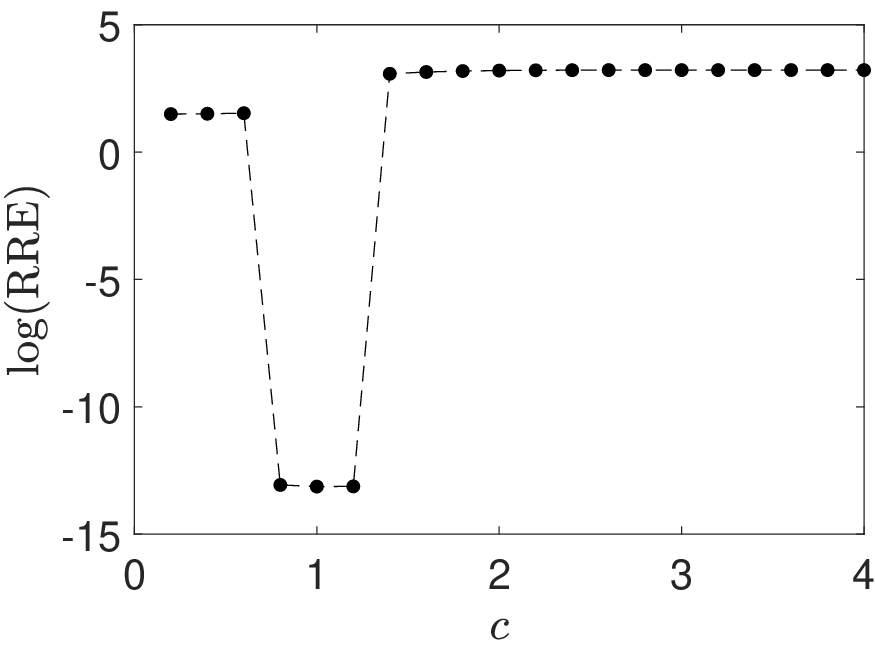}}\vskip 0pt
			\centerline{\scriptsize {(f)~$\alpha=0.5$}}\vskip -3pt
			\centerline{ }
		\end{minipage}
		
		\begin{minipage}{0.16\linewidth}
			\footnotesize
			\vspace{1pt}
			\centerline{\includegraphics[width=3cm]{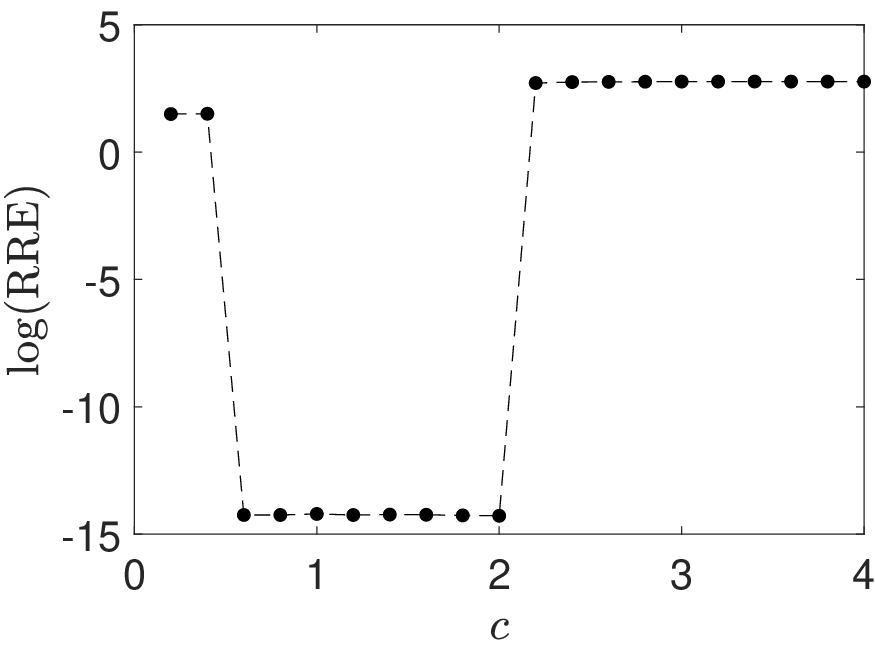}}\vskip 0pt
			\centerline{\scriptsize {(g)~$\beta=100$}}\vskip -3pt
			\centerline{ }
		\end{minipage}\hspace{16mm}
		\begin{minipage}{0.16\linewidth}
			\footnotesize
			\vspace{1pt}
			\centerline{\includegraphics[width=3cm]{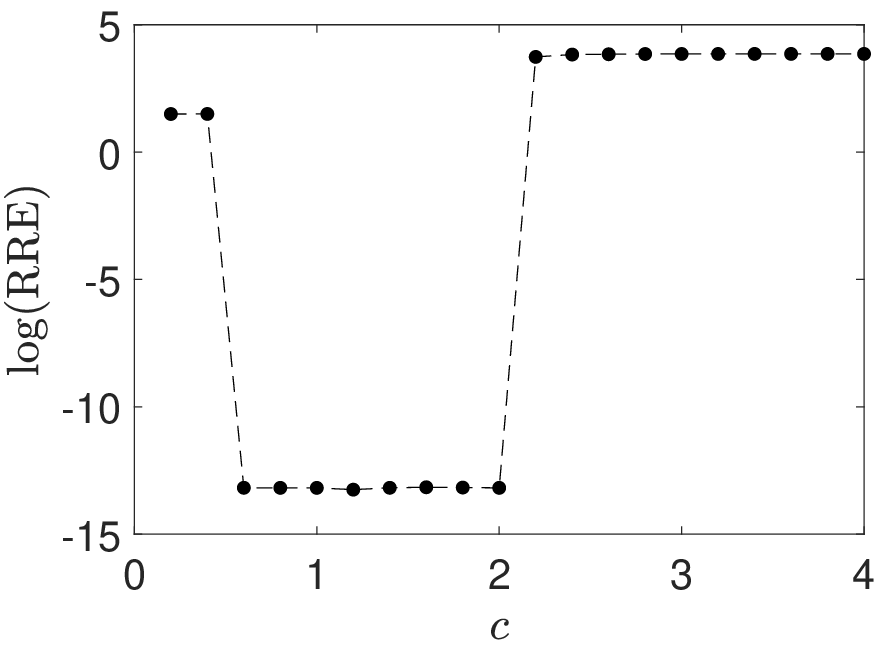}}\vskip 0pt
			\centerline{\scriptsize {(h)~$\beta=300$}}\vskip -3pt
			\centerline{ }
		\end{minipage}\hspace{16mm}
		\begin{minipage}{0.16\linewidth}
			\footnotesize
			\vspace{1pt}
			\centerline{\includegraphics[width=3cm]{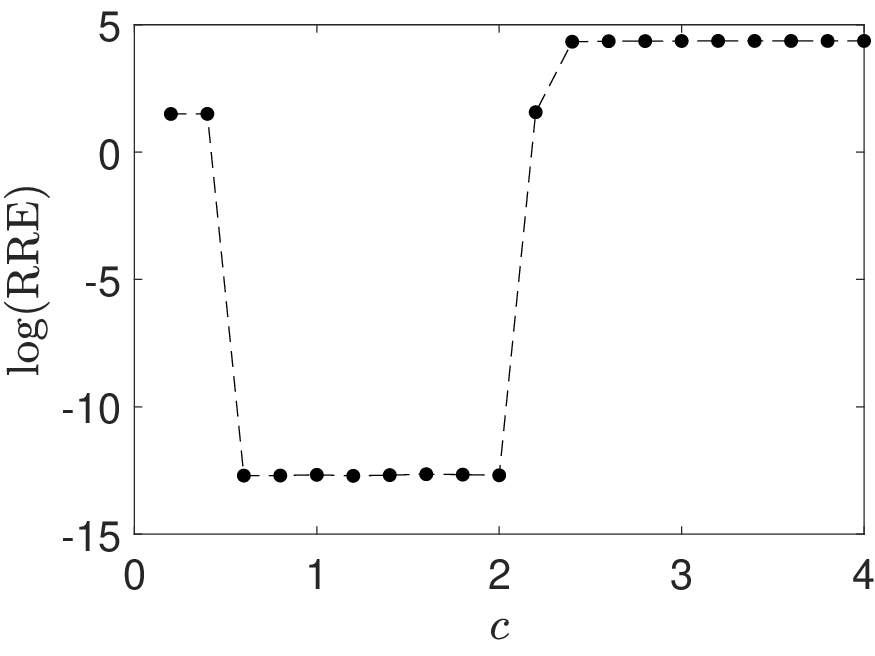}}\vskip 0pt
			\centerline{\scriptsize {(i)~$\beta=500$}}\vskip -3pt
			\centerline{ }
		\end{minipage}
		
		\begin{minipage}{0.16\linewidth}
			\footnotesize
			\vspace{1pt}
			\centerline{\includegraphics[width=3cm]{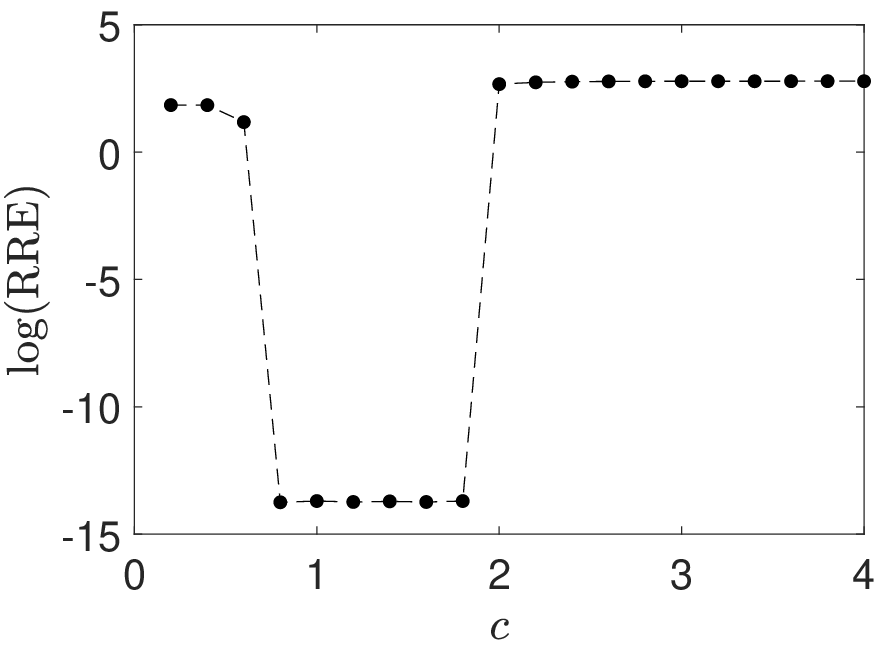}}\vskip 0pt
			\centerline{\scriptsize {(j)~$r = 40$}}\vskip -3pt
			\centerline{ }
		\end{minipage}\hspace{16mm}
		\begin{minipage}{0.16\linewidth}
			\footnotesize
			\vspace{1pt}
			\centerline{\includegraphics[width=3cm]{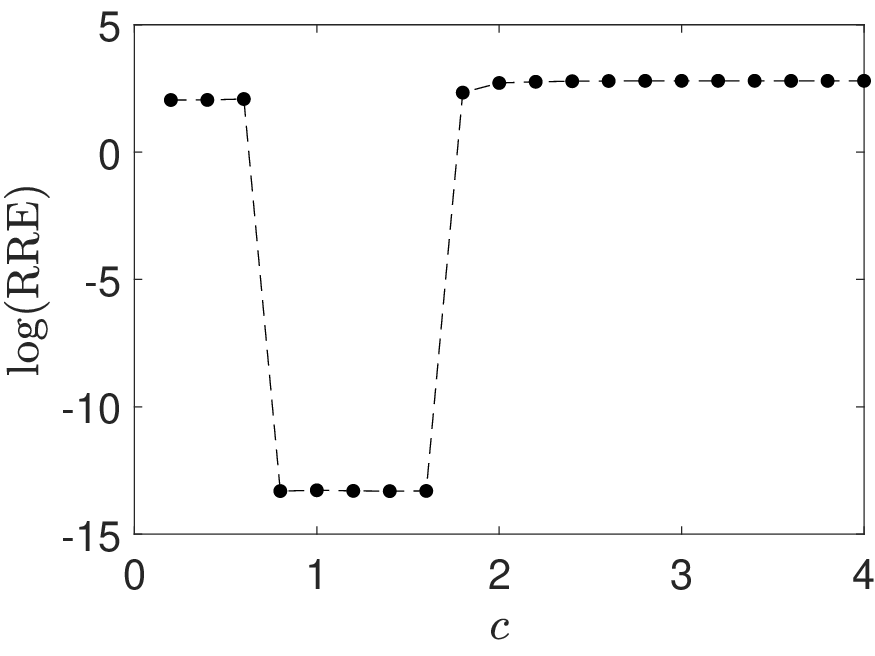}}\vskip 0pt
			\centerline{\scriptsize {(k)~$r = 60$}}\vskip -3pt
			\centerline{ }
		\end{minipage}\hspace{16mm}
		\begin{minipage}{0.16\linewidth}
			\footnotesize
			\vspace{1pt}
			\centerline{\includegraphics[width=3cm]{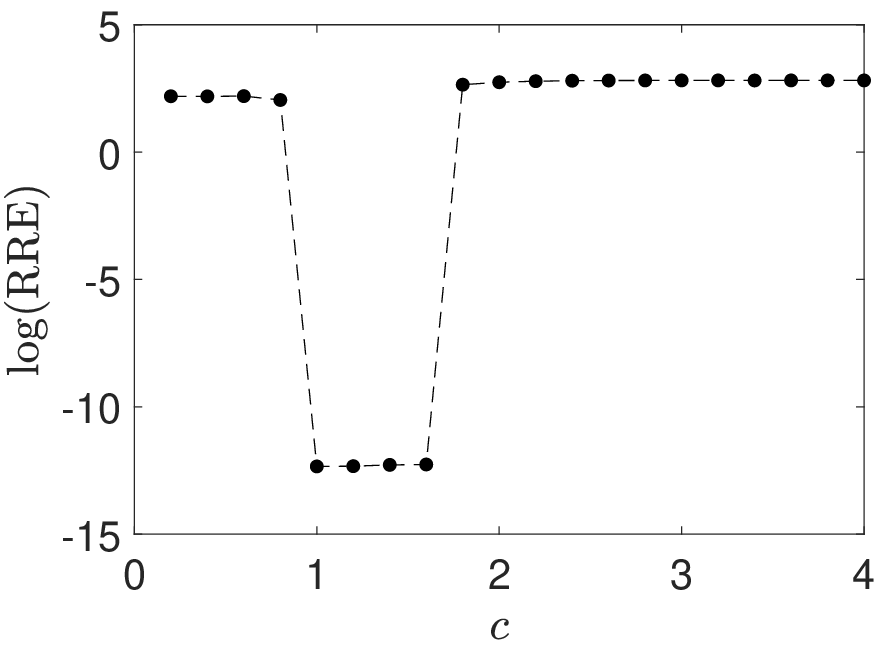}}\vskip 0pt
			\centerline{\scriptsize {(l)~$r = 80$}}\vskip -3pt
			\centerline{ }
		\end{minipage}
		\caption{Log-scale RRE versus $c$ where $\lambda = c/\sqrt{\max(m,n)}$. (a)-(c) plot the RRE versus $c$ for different percentage observation ratios $\gamma$ at $r=20$, $\alpha = 0.2$ and $\beta = 100$. (d)-(f) show the RRE versus $c$ for different outlier ratios $\alpha$ at $r=20$, $\gamma = 0.8$ and $\beta = 100$. (g)-(i) plot the RRE versus $c$ for different outlier maximum values $\beta$ at $r=20$, $\gamma = 0.8$ and $\alpha = 0.2$. (j)-(l) plot the RRE versus $c$ for different matrix ranks $r$ at $\beta = 100$, $\gamma = 0.8$ and $\alpha = 0.2$.}
		\label{lambda_choice}
	\end{figure}
	
	\begin{figure}
		\centering
		\begin{minipage}{0.23\linewidth}
			\footnotesize
			\vspace{1pt}
			\centerline{\includegraphics[width=4.2cm]{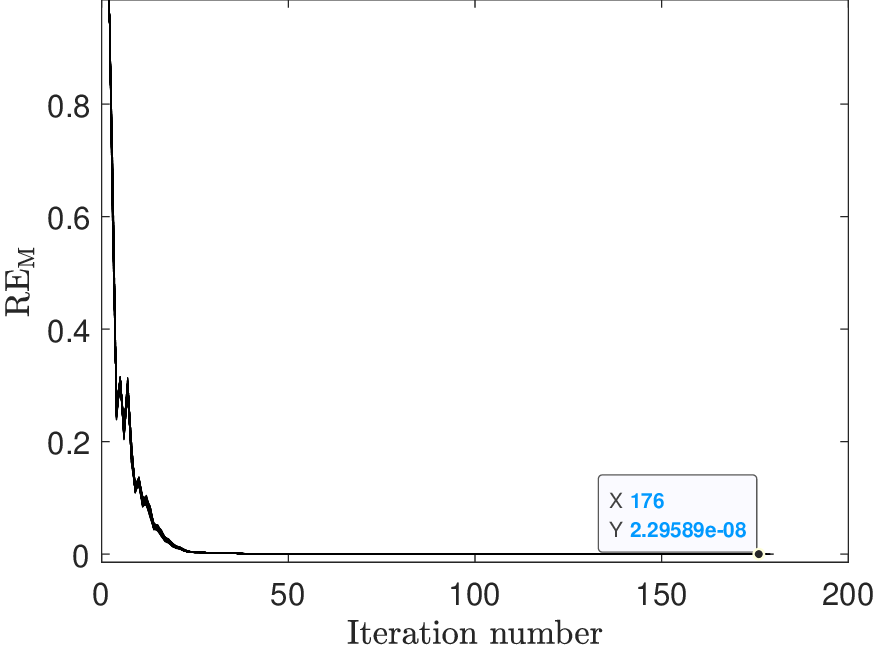}}\vskip 0pt
			\centerline{\scriptsize {(a)}}\vskip -3pt
			\centerline{ }
		\end{minipage}\hspace{23mm}
		\begin{minipage}{0.23\linewidth}
			\footnotesize
			\vspace{1pt}
			\centerline{\includegraphics[width=4.2cm]{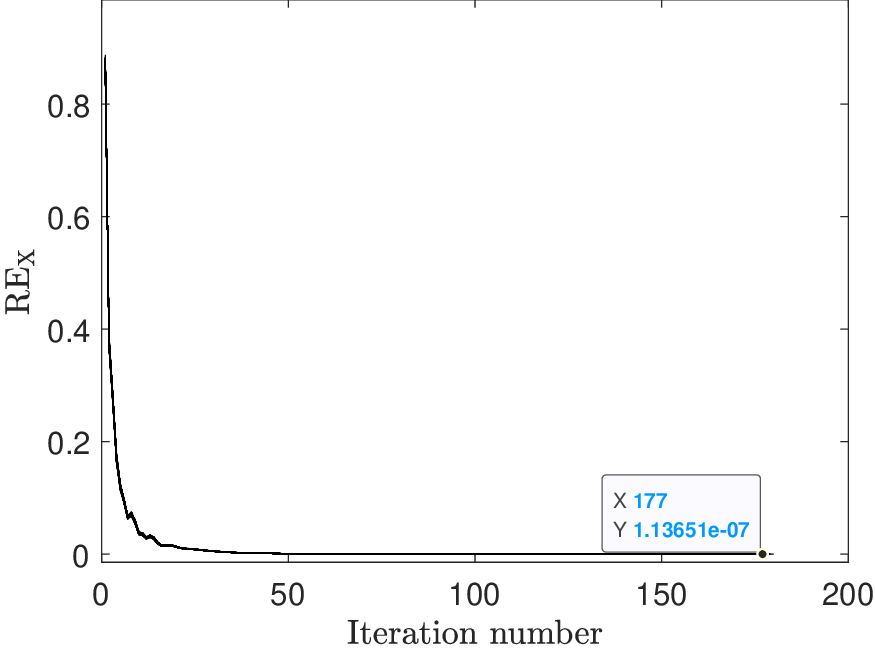}}\vskip 0pt
			\centerline{\scriptsize {(b)}}\vskip -3pt
			\centerline{ }
		\end{minipage}
		\caption{Convergence curves of the proposed algorithm. }
		\label{convergence_MX}
	\end{figure}
	We first conduct a series of experiments on the choice of the hyper-parameter $\lambda$ where $\lambda = c/\sqrt{\max(m,n)}$. We set $m=n=400$ for convenience. Fig.~\ref{lambda_choice} plots the RRE versus $\lambda$ for various parameters settings, including different observations, outlier levels and matrix ranks. Figs.~\ref{lambda_choice} (a)-(c) show the influence of the observation ratio $\gamma$ on the recovery error, and it is seen that there is a wide range for the choice of $\lambda$ even when $\gamma$ decreases. Figs. (d)-(f) and (g)-(i) show the impact of the outlier ratio $\alpha$ and  the outlier maximum magnitude $\beta$ on $\lambda$, respectively. We observe that the outlier magnitude has little influence on the choice of $\lambda$ because the proposed loss function is bounded from above, while the proper range of $\lambda$ becomes smaller when $\alpha$ increases. 
	Figs. (j)-(l) show the impact of the rank on $\lambda$, and it is observed that the admissible range of $\lambda$ decreases as the rank increases. 
	We set $\lambda=1/\sqrt{\max(m,n)}$ for convenience, because it attains comparable recovery results although $\lambda$ is not the optimal value for the current settings.
	
	In addition, the convergence of the developed algorithm is investigated. To this end, two evaluation metrics are adopted:
	\begin{equation}
		\begin{split}
			{\rm RE_{\pmb M^k}}&=\left\|{\pmb M}^k-\pmb M^{k-1}\right\|_F{ /}\left\|\pmb M^{k-1}\right\|_F\\
			{\rm RE_{\pmb X^k}}&=\left\|\pmb X-\pmb M^k-\pmb S^k\right\|_F/\left\|\pmb X\right\|_F
		\end{split}
	\end{equation}
	where $\pmb X$ is the observed matrix. Fig.~\ref{convergence_MX} shows the convergence curves of $100$ independent runs. It is seen that ${\rm RE_{\pmb M^k}}$ and ${\rm RE_{\pmb X^k}}$ approach zero when the algorithm converges.
	
\begin{figure}
		\centering
		\begin{minipage}{0.23\linewidth}
			\footnotesize
			\vspace{1pt}
			\centerline{\includegraphics[width=4.4cm]{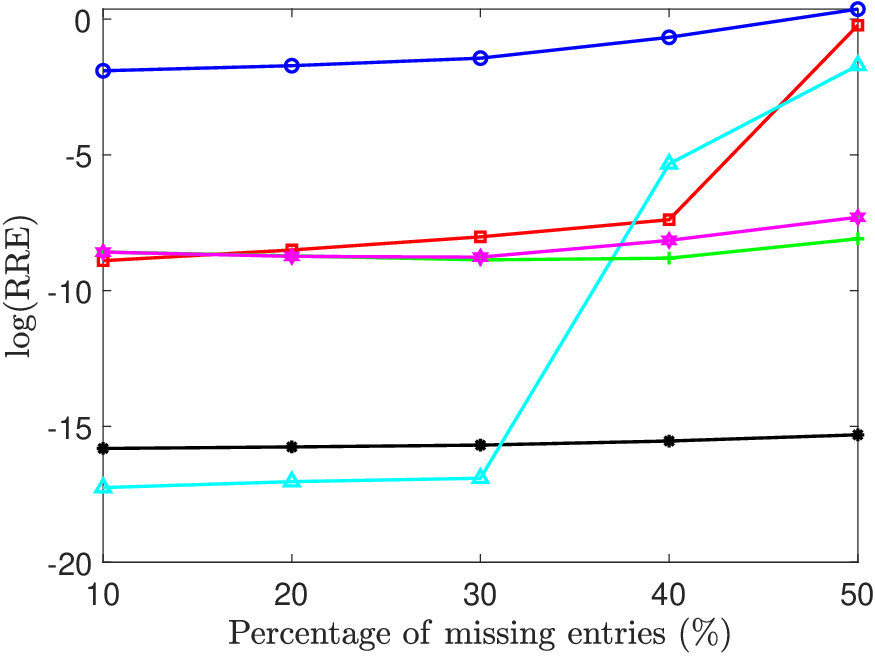}}\vskip 0pt
			\centerline{\scriptsize {(a)}}\vskip -3pt
			\centerline{ }
		\end{minipage}\hspace{23mm}
		\begin{minipage}{0.23\linewidth}
			\footnotesize
			\vspace{1pt}
			\centerline{\includegraphics[width=4.4cm]{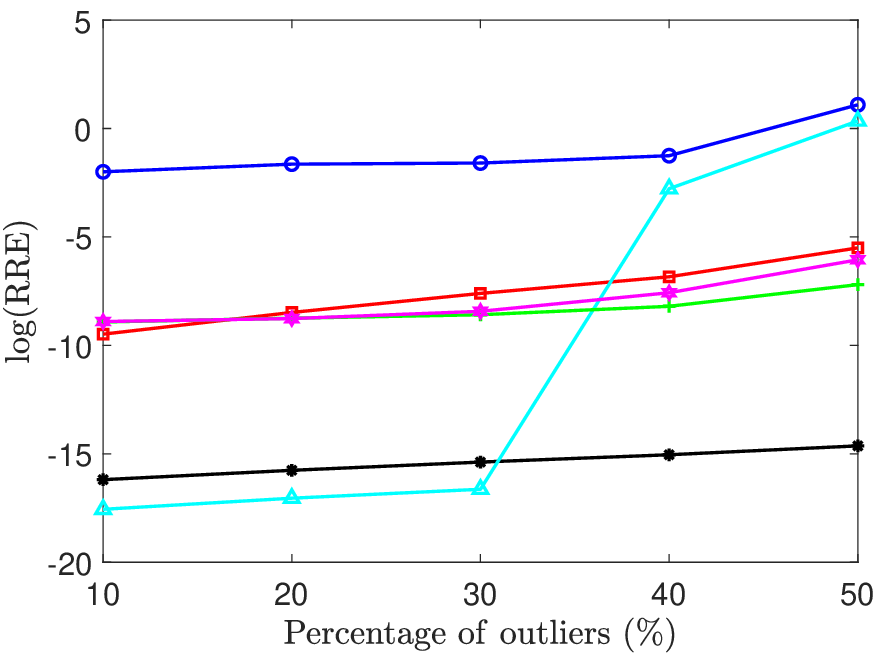}}\vskip 0pt
			\centerline{\scriptsize {(b)}}\vskip -3pt
			\centerline{ }
		\end{minipage}
		
		\begin{minipage}{0.23\linewidth}
			\footnotesize
			\vspace{1pt}
			\centerline{\includegraphics[width=4.4cm]{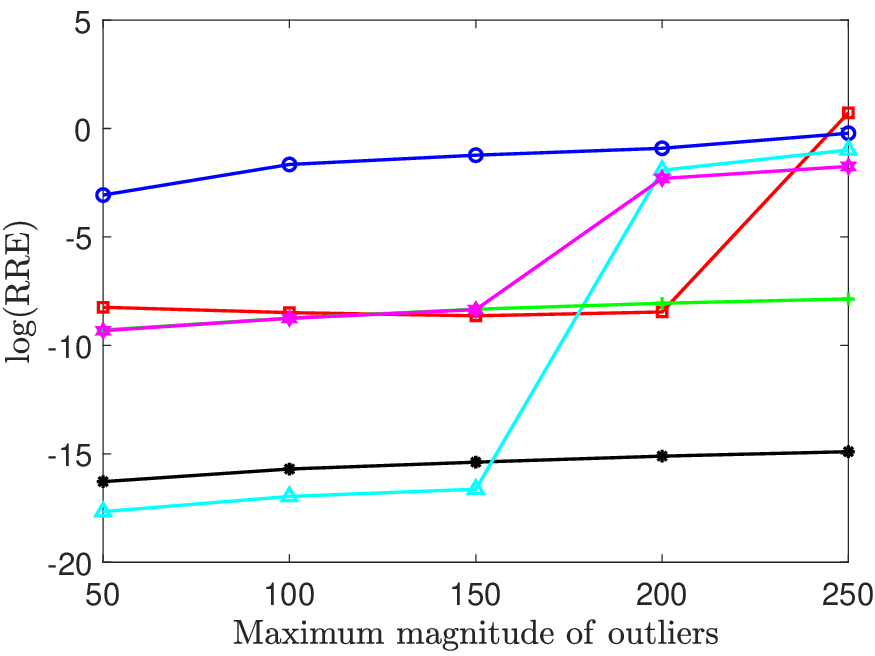}}\vskip 0pt
			\centerline{\scriptsize {(c)}}\vskip -3pt
			\centerline{ }
		\end{minipage}\hspace{23mm}
		\begin{minipage}{0.23\linewidth}
			\footnotesize
			\vspace{1pt}
			\centerline{\includegraphics[width=4.4cm]{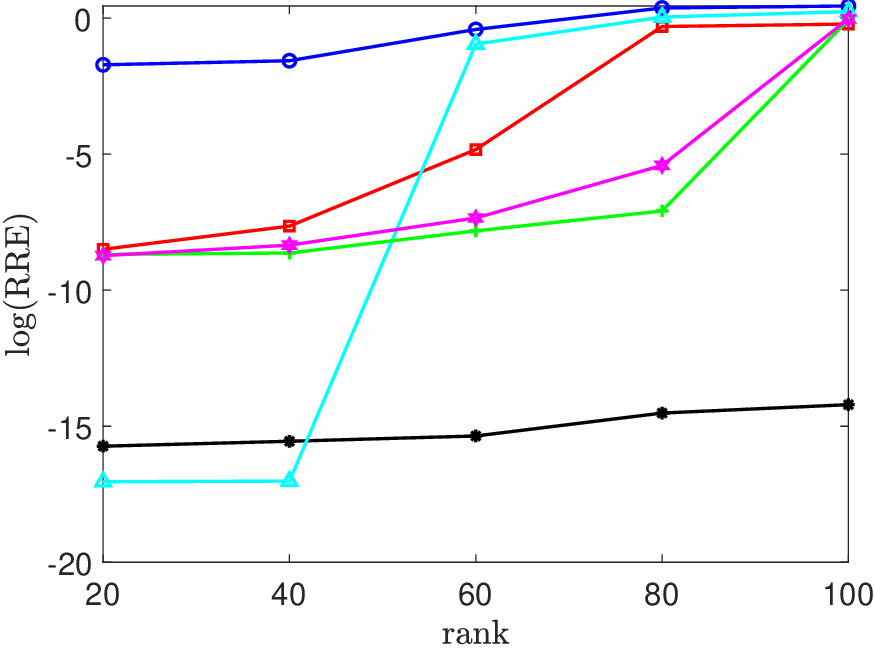}}\vskip 0pt
			\centerline{\scriptsize {(d)}}\vskip -3pt
			\centerline{ }
		\end{minipage}
		
		\begin{minipage}{0.45\linewidth}
			\footnotesize
			\vspace{1pt}
			\centerline{\includegraphics[width=8.8cm]{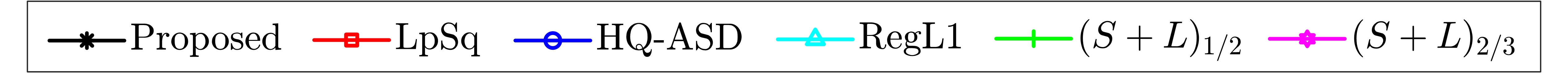}}\vskip 0pt
			\centerline{ }
		\end{minipage}
		\caption{Recovery performance for all algorithms under different scenarios. (a) RRE versus percentage of missing elements for different methods. (b) RRE versus percentage of outliers for different methods.(c) RRE versus outlier magnitude for different methods. (d) RRE versus matrix rank for different methods.}
		\label{RRE_cases}
	\end{figure}

	After choosing a proper value of $\lambda$, we compare our algorithm with the competitors for different cases. Fig.~\ref{RRE_cases} (a) plots the $\rm \log (RRE)$ curves versus the percentage of missing entries for different methods. We observe that compared with HQ-ASD, $(\pmb S$+$\pmb L)_{{1}/{2}}$, $(\pmb S$+$\pmb L)_{{2}/{3}}$ and $\rm LpSq$, $\rm RegL_1$ and our method have a better recovery performance when the percentage of missing entries is less than $30\%$. It is seen that the proposed algorithm outperforms all competing techniques for a higher missing ratio. Fig.~\ref{RRE_cases} (b) compares the recovery results under a varying percentage of outliers. Similarly, the NNSR is superior to the remaining approaches when the percentage of outliers is larger than $30\%$. Figs.~\ref{RRE_cases} (c) and (d) plot the $\rm \log (RRE)$ versus the magnitude of outliers and the matrix rank, respectively. Compared with the competitors, the proposed method has a stable recovery result for different outlier magnitudes and matrix ranks.

	\begin{figure}[htb]
		\centering
		\centerline{\scriptsize {Image-1~~~Image-2~~~Image-3~~~Image-4~~~Image-5~~~Image-6~~~Image-7~~~Image-8}}\vskip 1pt
		\includegraphics[width=8.5cm]{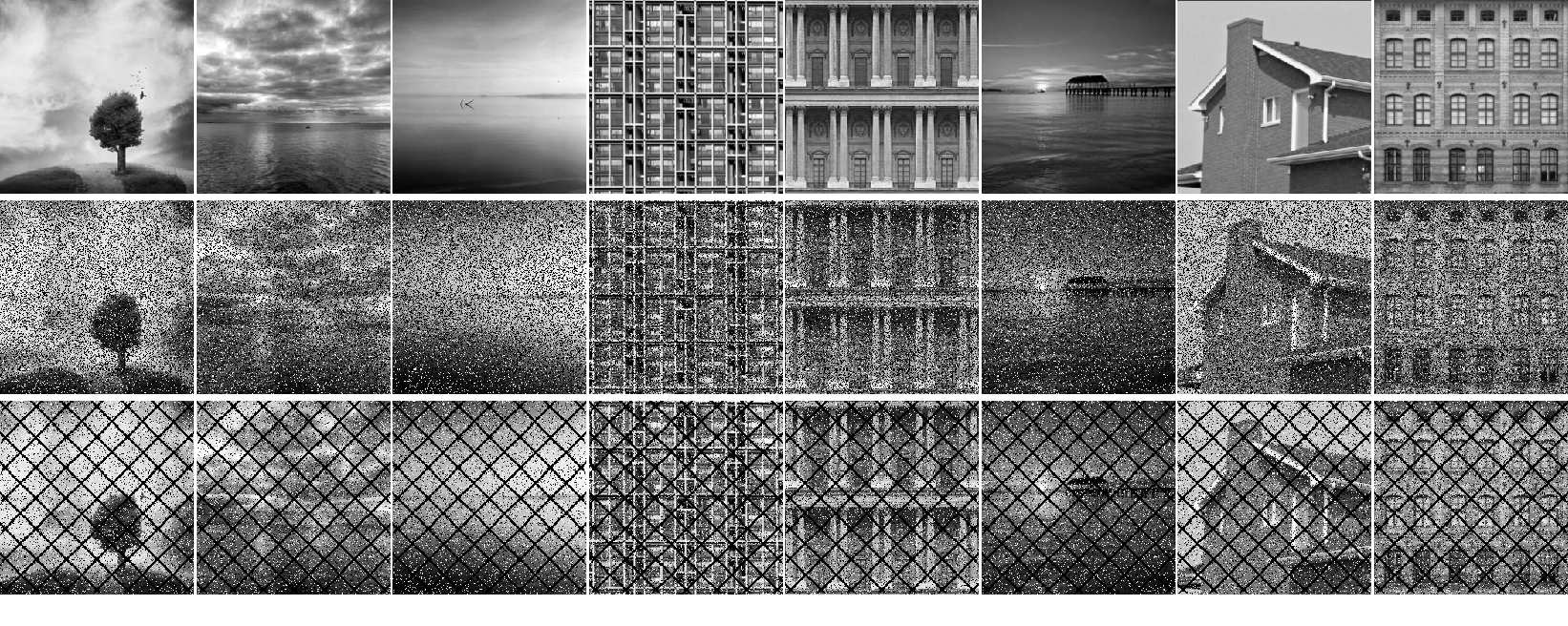}
		\vspace{-0.5em}
		\caption{Test images}\label{Eight_Test_images}
	\end{figure} 
	\begin{figure}
		\centering
		\begin{minipage}{0.23\linewidth}
			\footnotesize
			\vspace{1pt}
			\centerline{\includegraphics[width=8.5cm]{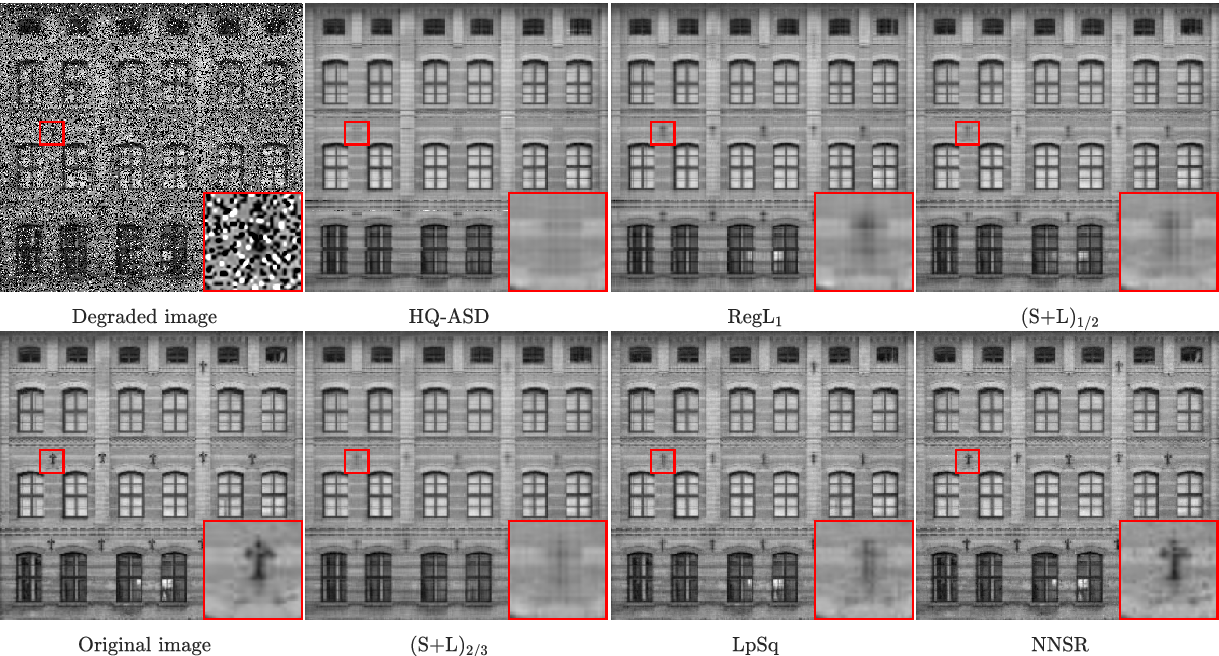}}\vskip 0pt
			\centerline{\scriptsize {(a)~Image restoration results for the random mask}}\vskip -3pt
			\centerline{ }
		\end{minipage}
		
		\begin{minipage}{0.45\linewidth}
			\footnotesize
			\vspace{1pt}
			\centerline{\includegraphics[width=8.5cm]{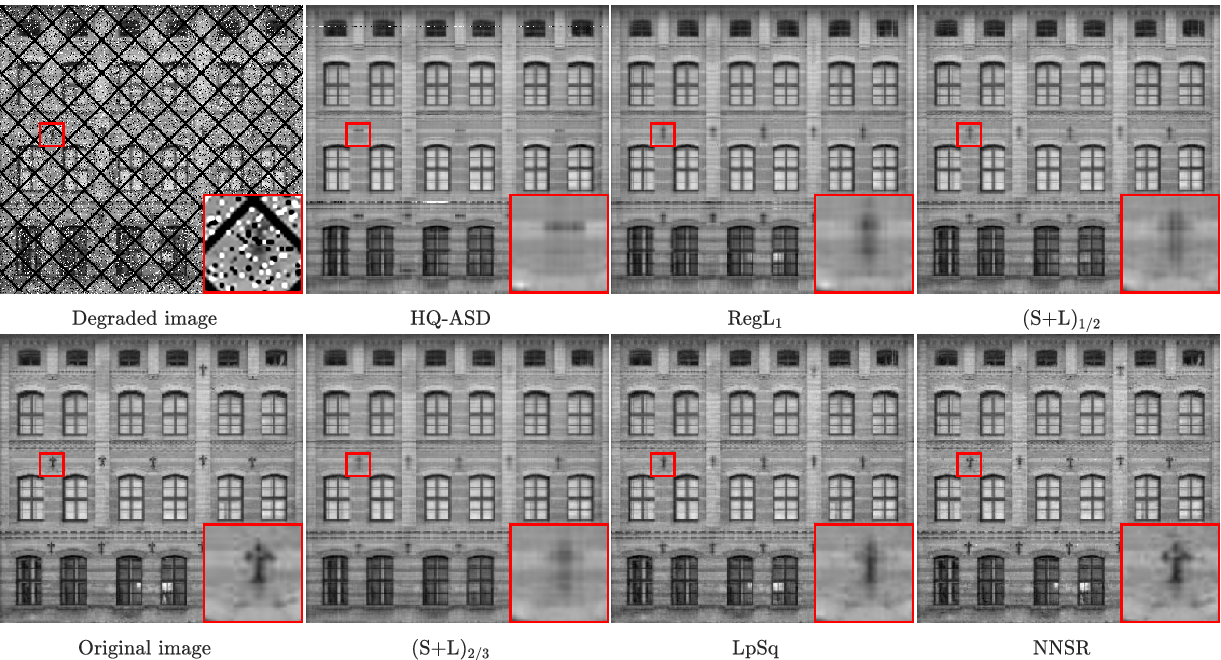}}\vskip 0pt
			\centerline{\scriptsize {(b)~Image restoration results for the stripe mask}}\vskip -3pt
			\centerline{ }
		\end{minipage}
		\caption{Image restoration results for Image-8}
		\label{Ima_res_8}
	\end{figure}

	\begin{table*}
		\caption{\small {Image restoration results from different algorithms in terms of PSNR and SSIM. The best and the second best results are highlighted in bold and underlined. The results are the average value of $20$ trials.}}  
		\begin{center}
			\setlength{\tabcolsep}{1.5mm}{
				\begin{tabular}{cccccccccccccc}
					\hline
					{\multirow{2}{*}{}} & \multicolumn {1} {c} {} & \multicolumn {6} {c} {PSNR} & \multicolumn {6} {c} {SSIM}\\
					\cmidrule(lr){3-8} \cmidrule(lr){9-14}
					\multicolumn{1}{c}{} &\multicolumn{1}{c}{} &\multicolumn{1}{c}{$\rm HQ$-$\rm ASD$} & \multicolumn{1}{c}{$\rm RegL_1$}  & \multicolumn{1}{c}{$\rm (S$+$\rm L)_{1/2}$}&\multicolumn{1}{c}{$\rm (S$+$\rm L)_{2/3}$}&\multicolumn{1}{c}{$\rm LpSq$} & \multicolumn{1}{c}{$\rm NNSR$} &\multicolumn{1}{c}{$\rm HQ$-$\rm ASD$} & \multicolumn{1}{c}{$\rm RegL_1$}  & \multicolumn{1}{c}{$\rm (S$+$\rm L)_{1/2}$}&\multicolumn{1}{c}{$\rm (S$+$\rm L)_{2/3}$} &\multicolumn{1}{c}{$\rm LpSq$} & \multicolumn{1}{c}{$\rm NNSR$}\\
					\hline
					\multirow{9}{*}{\begin{tabular}[c]{@{}c@{}}Random \\ mask\end{tabular}} & \multicolumn{1}{c}{Image-1} &\multicolumn{1}{c}{25.135} & \multicolumn{1}{c}{ 26.589} &\multicolumn{1}{c}{{ \underline {29.540}}} & \multicolumn{1}{c}{ 29.217} &\multicolumn{1}{c}{ 28.682}&\multicolumn{1}{c}{\bf 31.300} & \multicolumn{1}{c}{ 0.8179} &\multicolumn{1}{c}{{ 0.8207}} & \multicolumn{1}{c}{ 0.9000} &\multicolumn{1}{c}{  \underline{0.9016}}& \multicolumn{1}{c}{ 0.8938} &\multicolumn{1}{c}{\bf 0.9295}\\
					& \multicolumn{1}{c}{Image-2} &\multicolumn{1}{c}{26.868} & \multicolumn{1}{c}{ 26.908} &\multicolumn{1}{c}{{ 31.356}} & \multicolumn{1}{c}{ 31.366} &\multicolumn{1}{c}{ \underline {32.739}}&\multicolumn{1}{c}{\bf 34.990} & \multicolumn{1}{c}{ 0.7479} &\multicolumn{1}{c}{{ 0.7347}} & \multicolumn{1}{c}{ 0.8528} &\multicolumn{1}{c}{ 0.8559}& \multicolumn{1}{c}{  \underline{0.8908}} &\multicolumn{1}{c}{\bf 0.9305}\\
					& \multicolumn{1}{c}{Image-3} &\multicolumn{1}{c}{34.792} & \multicolumn{1}{c}{ 35.060} &\multicolumn{1}{c}{{ 39.608}} & \multicolumn{1}{c}{ \underline {40.235}} &\multicolumn{1}{c}{ 39.458}&\multicolumn{1}{c}{\bf 40.541} & \multicolumn{1}{c}{ 0.9396} &\multicolumn{1}{c}{{ 0.9544}} & \multicolumn{1}{c}{ 0.9771} &\multicolumn{1}{c}{  \underline{0.9836}}& \multicolumn{1}{c}{ 0.9763} &\multicolumn{1}{c}{\bf 0.9887}\\
					& \multicolumn{1}{c}{Image-4} &\multicolumn{1}{c}{20.848} & \multicolumn{1}{c}{ 23.702} &\multicolumn{1}{c}{{ 25.056}} & \multicolumn{1}{c}{ 24.515} &\multicolumn{1}{c}{  \underline{26.779}}&\multicolumn{1}{c}{\bf 26.955} & \multicolumn{1}{c}{ 0.8902} &\multicolumn{1}{c}{{ 0.9109}} & \multicolumn{1}{c}{ 0.9406} &\multicolumn{1}{c}{ 0.9375}& \multicolumn{1}{c}{\bf 0.9572} &\multicolumn{1}{c}{ \underline {0.9531}}\\
					& \multicolumn{1}{c}{Image-5} &\multicolumn{1}{c}{25.791} & \multicolumn{1}{c}{ 23.311} &\multicolumn{1}{c}{{ 28.697}} & \multicolumn{1}{c}{ 28.709} &\multicolumn{1}{c}{ \underline {28.966}}&\multicolumn{1}{c}{\bf 28.975} & \multicolumn{1}{c}{ 0.8508} &\multicolumn{1}{c}{{ 0.8271}} & \multicolumn{1}{c}{ 0.8945} &\multicolumn{1}{c}{ 0.8965}& \multicolumn{1}{c}{ \bf 0.9093} &\multicolumn{1}{c}{ \underline{0.8977}}\\
					& \multicolumn{1}{c}{Image-6} &\multicolumn{1}{c}{26.530} & \multicolumn{1}{c}{ 27.978} &\multicolumn{1}{c}{{ \underline {33.252}}} & \multicolumn{1}{c}{ 32.285} &\multicolumn{1}{c}{ 31.871}&\multicolumn{1}{c}{\bf 33.341} & \multicolumn{1}{c}{ 0.8004} &\multicolumn{1}{c}{{ 0.7987}} & \multicolumn{1}{c}{  \underline{0.9370}} &\multicolumn{1}{c}{ 0.9365}& \multicolumn{1}{c}{ 0.9242} &\multicolumn{1}{c}{\bf 0.9557}\\
					& \multicolumn{1}{c}{Image-7} &\multicolumn{1}{c}{19.857} & \multicolumn{1}{c}{ 22.538} &\multicolumn{1}{c}{{ \underline {25.623}}} & \multicolumn{1}{c}{24.254} &\multicolumn{1}{c}{ 22.964}&\multicolumn{1}{c}{\bf 29.139} & \multicolumn{1}{c}{ 0.6617} &\multicolumn{1}{c}{{ 0.6963}} & \multicolumn{1}{c}{  \underline{0.8088}} &\multicolumn{1}{c}{ 0.8055}& \multicolumn{1}{c}{ 0.7752} &\multicolumn{1}{c}{\bf 0.8146}\\
					& \multicolumn{1}{c}{Image-8} &\multicolumn{1}{c}{26.123} & \multicolumn{1}{c}{ 27.192} &\multicolumn{1}{c}{{ 29.784}} & \multicolumn{1}{c}{ 29.718} &\multicolumn{1}{c}{ \underline {32.497}}&\multicolumn{1}{c}{\bf 33.596} & \multicolumn{1}{c}{ 0.8529} &\multicolumn{1}{c}{{ 0.8467}} & \multicolumn{1}{c}{ 0.8974} &\multicolumn{1}{c}{ 0.9005}& \multicolumn{1}{c}{  \underline{0.9387}} &\multicolumn{1}{c}{\bf 0.9422}\\
					\cmidrule(lr){2-14}
					& \multicolumn{1}{c}{Average} &\multicolumn{1}{c}{25.743} & \multicolumn{1}{c}{ 26.660} &\multicolumn{1}{c}{{ 30.365}} & \multicolumn{1}{c}{ 30.037} &\multicolumn{1}{c}{ \underline{30.495}}&\multicolumn{1}{c}{\bf 32.355} & \multicolumn{1}{c}{ 0.8202} &\multicolumn{1}{c}{{ 0.8237}} & \multicolumn{1}{c}{ 0.9010} &\multicolumn{1}{c}{ 0.9022}& \multicolumn{1}{c}{ \underline{0.9082}} &\multicolumn{1}{c}{\bf 0.9265}\\
					\hline
					\multirow{9}{*}{\begin{tabular}[c]{@{}c@{}}Fixed \\ mask\end{tabular}} & \multicolumn{1}{c}{Image-1} &\multicolumn{1}{c}{25.256} & \multicolumn{1}{c}{ 25.758} &\multicolumn{1}{c}{{  \underline{29.663}}} & \multicolumn{1}{c}{ 29.226} &\multicolumn{1}{c}{ 28.794}&\multicolumn{1}{c}{\bf 31.183} & \multicolumn{1}{c}{ 0.8185} &\multicolumn{1}{c}{{ 0.8135}} & \multicolumn{1}{c}{ 0.8950} &\multicolumn{1}{c}{  \underline{0.8973}}& \multicolumn{1}{c}{ 0.8882} &\multicolumn{1}{c}{\bf 0.9179}\\
					& \multicolumn{1}{c}{Image-2} &\multicolumn{1}{c}{26.858} & \multicolumn{1}{c}{ 26.392} &\multicolumn{1}{c}{{ 30.925}} & \multicolumn{1}{c}{ 30.979} &\multicolumn{1}{c}{  \underline{31.986}}&\multicolumn{1}{c}{\bf 33.568} & \multicolumn{1}{c}{ 0.7420} &\multicolumn{1}{c}{{ 0.7385}} & \multicolumn{1}{c}{ 0.8445} &\multicolumn{1}{c}{ 0.8482}& \multicolumn{1}{c}{  \underline{0.8760}} &\multicolumn{1}{c}{\bf 0.9140}\\
					& \multicolumn{1}{c}{Image-3} &\multicolumn{1}{c}{34.779} & \multicolumn{1}{c}{ 35.049} &\multicolumn{1}{c}{{ 39.653}} & \multicolumn{1}{c}{  \underline{40.222}} &\multicolumn{1}{c}{ 39.498}&\multicolumn{1}{c}{\bf 40.382} & \multicolumn{1}{c}{ 0.9416} &\multicolumn{1}{c}{{ 0.9548}} & \multicolumn{1}{c}{ 0.9774} &\multicolumn{1}{c}{  \underline{0.9831}}& \multicolumn{1}{c}{ 0.9767} &\multicolumn{1}{c}{\bf 0.9869}\\
					& \multicolumn{1}{c}{Image-4} &\multicolumn{1}{c}{20.827} & \multicolumn{1}{c}{ 22.793} &\multicolumn{1}{c}{{ 25.025}} & \multicolumn{1}{c}{ 24.578} &\multicolumn{1}{c}{\bf 26.650}&\multicolumn{1}{c}{  \underline{26.227}} & \multicolumn{1}{c}{ 0.8858} &\multicolumn{1}{c}{{ 0.9017}} & \multicolumn{1}{c}{ 0.9389} &\multicolumn{1}{c}{ 0.9363}& \multicolumn{1}{c}{\bf 0.9548} &\multicolumn{1}{c}{  \underline{0.9488}}\\
					& \multicolumn{1}{c}{Image-5} &\multicolumn{1}{c}{26.469} & \multicolumn{1}{c}{ 25.924} &\multicolumn{1}{c}{{ 28.577}} & \multicolumn{1}{c}{ 28.584} &\multicolumn{1}{c}{\bf 28.812}&\multicolumn{1}{c}{  \underline{28.717}} & \multicolumn{1}{c}{ 0.8492} &\multicolumn{1}{c}{{ 0.8397}} & \multicolumn{1}{c}{ 0.8926} &\multicolumn{1}{c}{ 0.8944}& \multicolumn{1}{c}{\bf 0.9066} &\multicolumn{1}{c}{  \underline{0.8973}}\\
					& \multicolumn{1}{c}{Image-6} &\multicolumn{1}{c}{26.307} & \multicolumn{1}{c}{ 26.621} &\multicolumn{1}{c}{{\bf 33.137}} & \multicolumn{1}{c}{ 32.452} &\multicolumn{1}{c}{ 31.797}&\multicolumn{1}{c}{  \underline{33.033}} & \multicolumn{1}{c}{ 0.7907} &\multicolumn{1}{c}{{ 0.7817}} & \multicolumn{1}{c}{ \underline{0.9352}} &\multicolumn{1}{c}{  {0.9351}}& \multicolumn{1}{c}{ 0.9221} &\multicolumn{1}{c}{\bf 0.9514}\\
					& \multicolumn{1}{c}{Image-7} &\multicolumn{1}{c}{19.825} & \multicolumn{1}{c}{ 22.657} &\multicolumn{1}{c}{{  \underline{24.822}}} & \multicolumn{1}{c}{24.159} &\multicolumn{1}{c}{ 22.945}&\multicolumn{1}{c}{\bf 27.532} & \multicolumn{1}{c}{ 0.6651} &\multicolumn{1}{c}{{ 0.6949}} & \multicolumn{1}{c}{ 0.7962} &\multicolumn{1}{c}{ \underline{0.7965}}& \multicolumn{1}{c}{ 0.7685} &\multicolumn{1}{c}{\bf 0.8034}\\
					& \multicolumn{1}{c}{Image-8}  & \multicolumn{1}{c}{ 25.647} &\multicolumn{1}{c}{{ 26.970}} & \multicolumn{1}{c}{ 29.929}&\multicolumn{1}{c}{29.705} &\multicolumn{1}{c}{  \underline{32.086}}&\multicolumn{1}{c}{\bf 32.528} & \multicolumn{1}{c}{ 0.8494} &\multicolumn{1}{c}{{ 0.8382}} & \multicolumn{1}{c}{ 0.8933} &\multicolumn{1}{c}{ 0.8964}& \multicolumn{1}{c}{ \underline{0.9303}} &\multicolumn{1}{c}{\bf 0.9330}\\
					\cmidrule(lr){2-14}
					& \multicolumn{1}{c}{Average} &\multicolumn{1}{c}{25.746} & \multicolumn{1}{c}{ 26.521} &\multicolumn{1}{c}{{ 30.216}} & \multicolumn{1}{c}{ 29.988} &\multicolumn{1}{c}{ \underline{30.321}}&\multicolumn{1}{c}{\bf 31.646} & \multicolumn{1}{c}{ 0.8178} &\multicolumn{1}{c}{{ 0.8204}} & \multicolumn{1}{c}{ 0.8966} &\multicolumn{1}{c}{ 0.8984}& \multicolumn{1}{c}{ \underline{0.9029}} &\multicolumn{1}{c}{\bf 0.9191}\\
					\hline
			\end{tabular}}
			\vspace{-2em}
			\label{twomask_results}
		\end{center}
	\end{table*}

	\subsection{Image Inpainting}
	Numerous real-world natural images can be approximated by low-rank matrices since their main information lies in a low-dimensional subspace~\cite{LinZhou-2018,ShangFH-2018}. Fig.~\ref{Eight_Test_images} shows several images in~\cite{HuY2013} and~\cite{Rtensor2023}, which are used to test the algorithms. These images may be incomplete due to a photosensitive device or shadow cast, and be contaminated by outliers during a wireless transmission or the acquisition stage. 
	In this section, the images are first converted into grayscale images and $20\%$ entries are corrupted by outliers with magnitudes in the range of $[-2,2]$. Besides, two different masks, that is, a random mask and a fixed mask, are utilized to cover the original images. A random mask selects the missing pixels randomly, while a fixed mask is the deterministic stripe in this study. In Fig.~\ref{Eight_Test_images}, the first row contains the true images, while the second and the third rows are degraded images covered by the random and stripe masks, respectively. Moreover, to measure the recovery performance, the peak signal-to-noise ratio (PSNR) and the structural index similarity (SSIM), are adopted, and the built-in commands `${\rm psnr (recovered, original)}$' and `${\rm ssim (recovered, original)}$' in MATLAB are employed to calculate them. Note that the competitors, such as HQ-ASD and $\rm RegL_1$, which are based on matrix factorization, require the matrix rank. Similar to~\cite{NieFrobust2013}, the rank $r$ is varied in the set $\{1,2,\cdots,30\}$, and its value is determined based on the highest PSNR value.
	
	Table~\ref{twomask_results} shows the restoration results for different algorithms. It is seen that when images are covered by a random mask, the proposed algorithm has the best recovery performance in terms of PSNR and has the highest average SSIM value, although its SSIM is inferior to LpSq for two images. Again, for the fixed mask, compared with HQ-ASD, $\rm RegL_1$, $(\pmb S$+$\pmb L)_{{1}/{2}}$, $(\pmb S$+$\pmb L)_{{2}/{3}}$ and $\rm LpSq$, NNSR achieves the best average restoration in terms of PSNR and SSIM. In addition, Fig.~\ref{Ima_res_8} shows the recovery results of Image-8 for different algorithms. We easily observe that NNSR gives a clearer visual result compared to the remaining methods.

	
	\subsection{Multispectral Imaging Restoration}
	Multispectral imaging (MSI) acquires images of the same scene using different wavelengths, and has numerous applications such as documents and artworks. However, these images may be contaminated by impulsive noise and suffer data loss due to photon effects and calibration errors. Thus, there is a need to improve the MSI quality. Two datasets from CAVE~\cite{CAVEdataset}, namely, feathers and flowers, are employed to evaluate the algorithms. Each dataset contains $31$ spectral bands with dimensions $512\times 512$. The data matrix $\pmb X \in \mathbb{R}^{262144\times 31}$ is constructed by vectorizing each band. Besides, $20\%$ of pixels in $\pmb X$ are randomly removed, and $10~{\rm dB}$ salt-and-pepper noise produced by the built-in MATLAB function `${\rm imnoise(\pmb I, salt~ \&~ pepper, \rho)}$' is added to the incomplete matrix. The relationship between $\rho$ and the signal-to-noise ratio (SNR) is $\rho = 1/{\rm SNR}$. 
	
	\begin{table}
		\caption{\small {MSI restoration results from different algorithms in terms of PSNR, SSIM and runtime. The best and second best results for each row are highlighted in bold and underlined. The results are the average value of $20$ independent runs.}}  
		\begin{center}
			\setlength{\tabcolsep}{0.6mm}{
				\begin{tabular}{cccccccc}
					\hline
					\multicolumn{1}{c}{} &\multicolumn{1}{c}{} &\multicolumn{1}{c}{$\rm HQ$-$\rm ASD$} & \multicolumn{1}{c}{$\rm RegL_1$}  & \multicolumn{1}{c}{$\rm (S$+$\rm L)_{1/2}$}&\multicolumn{1}{c}{$\rm (S$+$\rm L)_{2/3}$}&\multicolumn{1}{c}{$\rm LpSq$} & \multicolumn{1}{c}{$\rm NNSR$}\\
					\hline
					\multirow{3}{*}{\begin{tabular}[c]{@{}c@{}}feathers\end{tabular}} & \multicolumn{1}{c}{PSNR} &\multicolumn{1}{c}{24.978} & \multicolumn{1}{c}{ 32.304} &\multicolumn{1}{c}{{ \underline {41.124}}} & \multicolumn{1}{c}{ 33.751} &\multicolumn{1}{c}{ 40.134}&\multicolumn{1}{c}{\bf 43.495} \\
					& \multicolumn{1}{c}{SSIM} &\multicolumn{1}{c}{0.4931} & \multicolumn{1}{c}{ 0.8656} &\multicolumn{1}{c}{{ \underline {0.9539}}} & \multicolumn{1}{c}{ 0.9521} &\multicolumn{1}{c}{ 0.9489}&\multicolumn{1}{c}{\bf 0.9675} \\
					& \multicolumn{1}{c}{Runtime} &\multicolumn{1}{c}{264.96} & \multicolumn{1}{c}{ \bf 37.007} &\multicolumn{1}{c}{{ {162.23}}} & \multicolumn{1}{c}{ 144.44} &\multicolumn{1}{c}{ 1145.4}&\multicolumn{1}{c}{\underline {103.10}} \\
					\hline
					\multirow{3}{*}{\begin{tabular}[c]{@{}c@{}}flowers\end{tabular}} & \multicolumn{1}{c}{PSNR} &\multicolumn{1}{c}{26.169} & \multicolumn{1}{c}{ 34.228} &\multicolumn{1}{c}{{ \underline {43.625}}} & \multicolumn{1}{c}{ 32.264} &\multicolumn{1}{c}{ 42.490}&\multicolumn{1}{c}{\bf 46.437} \\
					& \multicolumn{1}{c}{SSIM} &\multicolumn{1}{c}{0.4678} & \multicolumn{1}{c}{ 0.8532} &\multicolumn{1}{c}{{ \underline {0.9685}}} & \multicolumn{1}{c}{ 0.9546} &\multicolumn{1}{c}{ 0.9603}&\multicolumn{1}{c}{\bf 0.9759} \\
					& \multicolumn{1}{c}{Runtime} &\multicolumn{1}{c}{252.30} & \multicolumn{1}{c}{ \bf{34.481}} &\multicolumn{1}{c}{{  {150.67}}} & \multicolumn{1}{c}{ 135.97} &\multicolumn{1}{c}{ 1121.7}&\multicolumn{1}{c}{ \underline{91.676}} \\
					\hline
			\end{tabular}}
			\vspace{-2em}
			\label{three_metrics}
		\end{center}
	\end{table}

	\begin{figure}
		\centering
		\begin{minipage}{0.23\linewidth}
			\footnotesize
			\vspace{1pt}
			\centerline{\includegraphics[width=4.2cm]{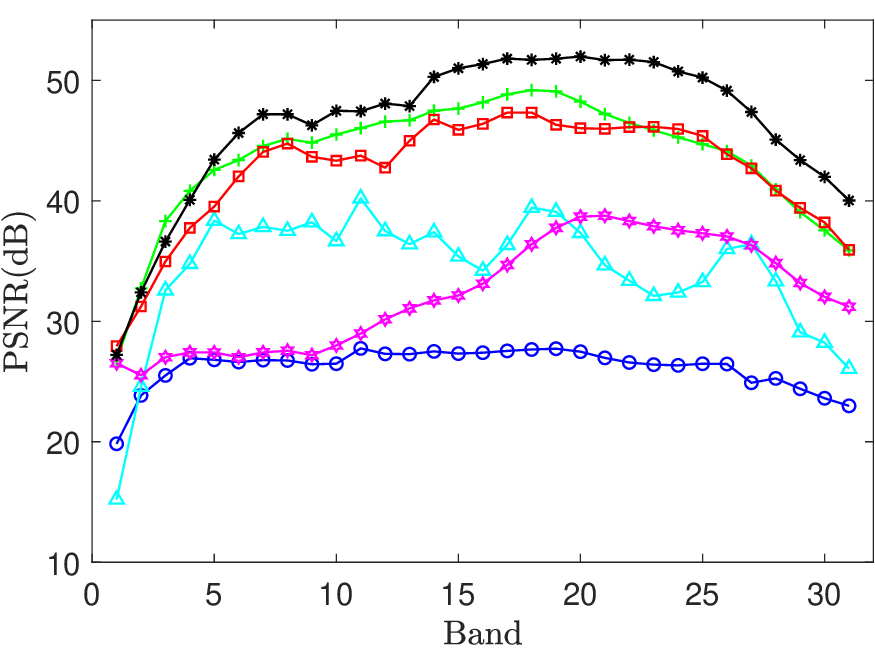}}\vskip 0pt
			\centerline{\scriptsize {(a)}}\vskip -3pt
			\centerline{ }
		\end{minipage}\hspace{23mm}
		\begin{minipage}{0.23\linewidth}
			\footnotesize
			\vspace{1pt}
			\centerline{\includegraphics[width=4.2cm]{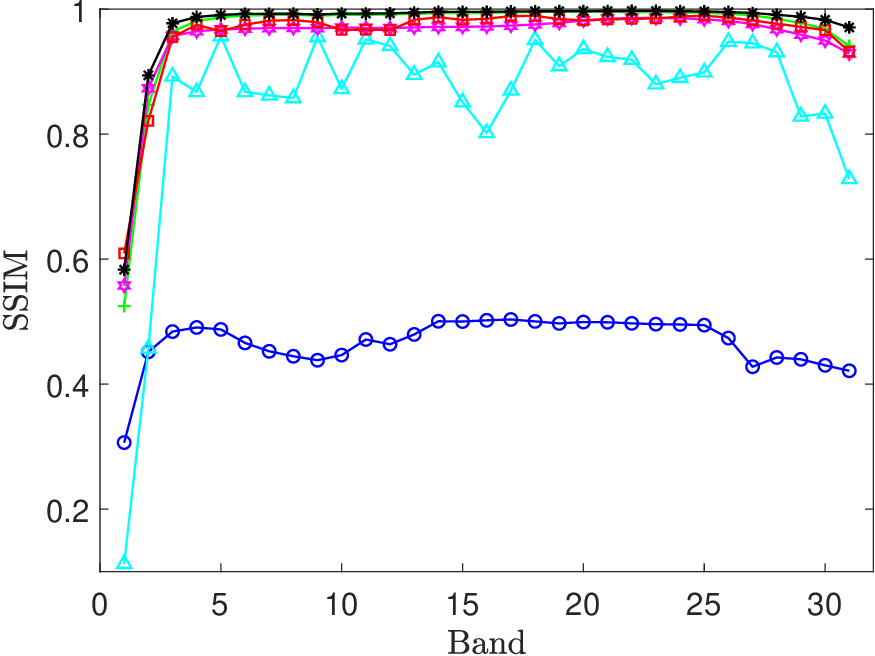}}\vskip 0pt
			\centerline{\scriptsize {(b)}}\vskip -3pt
			\centerline{ }
		\end{minipage}
		
		\begin{minipage}{0.45\linewidth}
			\footnotesize
			\vspace{1pt}
			\centerline{\includegraphics[width=8.8cm]{Legend_2.jpg}}\vskip 0pt
			\centerline{ }
		\end{minipage}
		\caption{Recovery performance for each band of `feathers' data in terms of PSNR and SSIM.}
		\label{band_feather}
	\end{figure}
	
	\begin{figure}
		\centering
		\begin{minipage}{0.23\linewidth}
			\footnotesize
			\vspace{1pt}
			\centerline{\includegraphics[width=4.2cm]{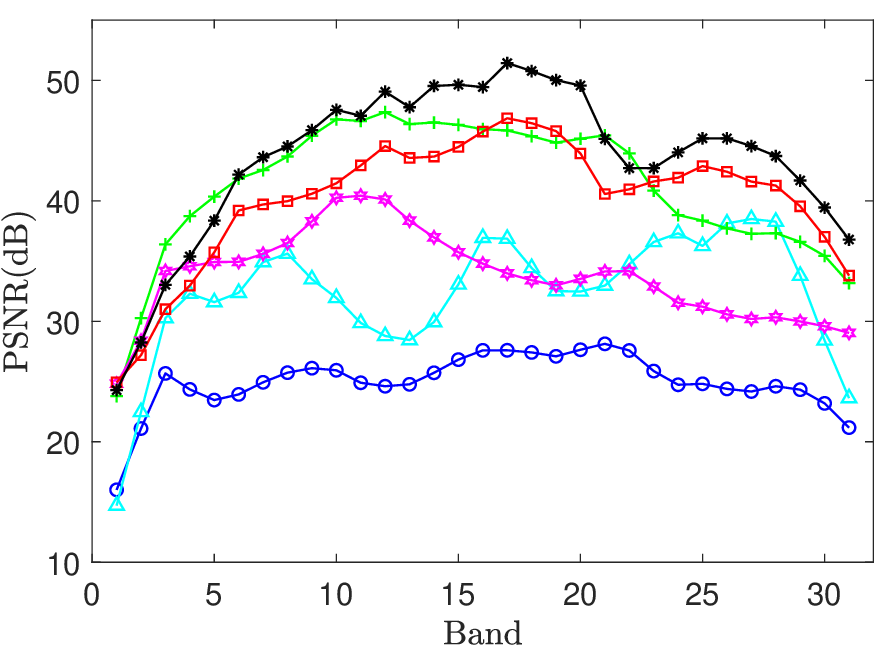}}\vskip 0pt
			\centerline{ }
		\end{minipage}\hspace{23mm}
		\begin{minipage}{0.23\linewidth}
			\footnotesize
			\vspace{1pt}
			\centerline{\includegraphics[width=4.2cm]{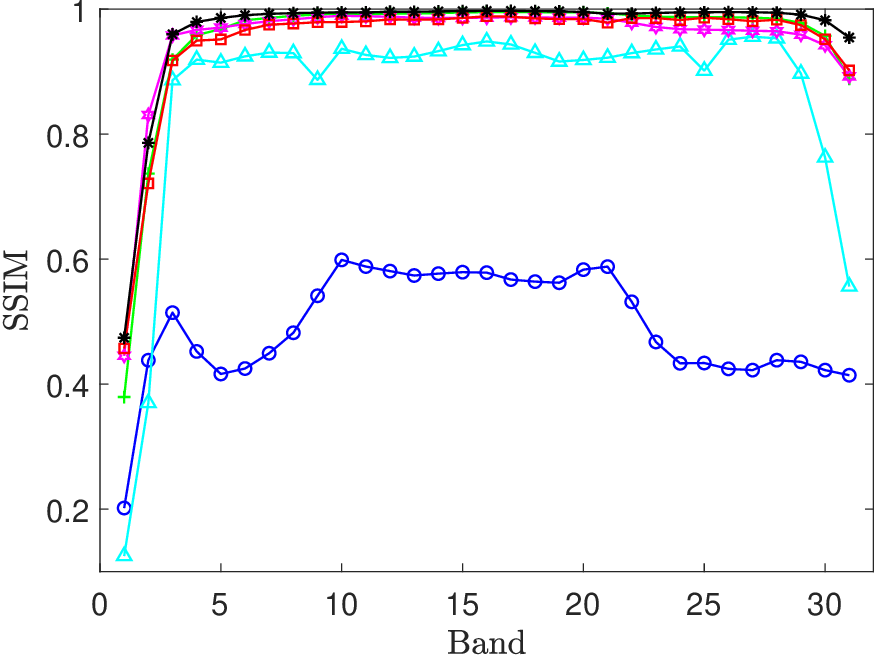}}\vskip 0pt
			\centerline{ }
		\end{minipage}
		
		\begin{minipage}{0.45\linewidth}
			\footnotesize
			\vspace{1pt}
			\centerline{\includegraphics[width=8.8cm]{Legend_2.jpg}}\vskip 0pt
			\centerline{ }
		\end{minipage}
		\caption{Recovery performance for each band of `flowers' data in terms of PSNR and SSIM.}
		\label{band_flower}
	\end{figure}

	\begin{figure}[htb]
		\centering
		\includegraphics[width=9cm]{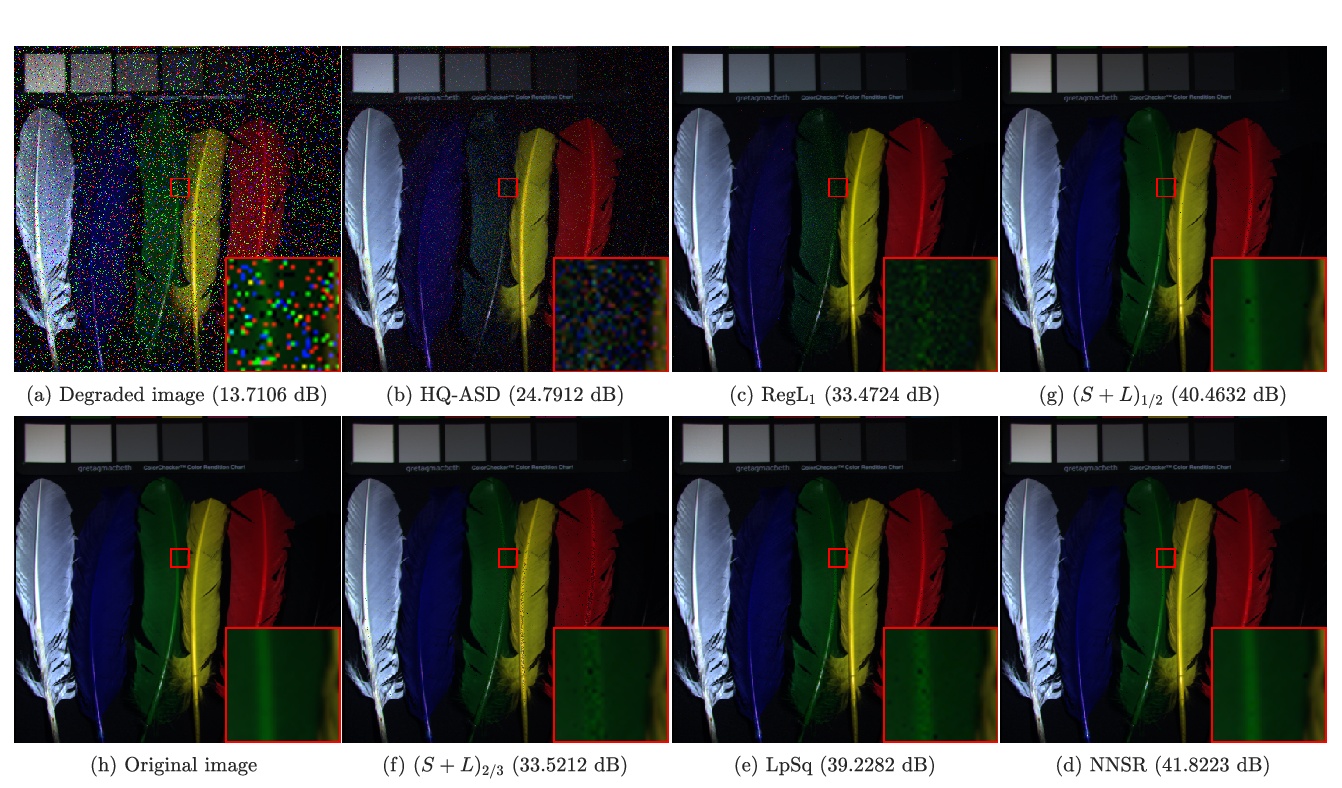}
		\vspace{-0.5em}
		\caption{Recovered images of `feathers' with bands 23-13-4 as R-G-B. (a) is the degraded image corrupted by impulsive noise and random mask, (h) is the original noise-free image, and the remaining images are the restoration results using different algorithms, with a demarcated area zoomed-in $6$ times.}\label{feather_images}
	\end{figure}

	\begin{figure}[htb]
		\centering
		\includegraphics[width=9cm]{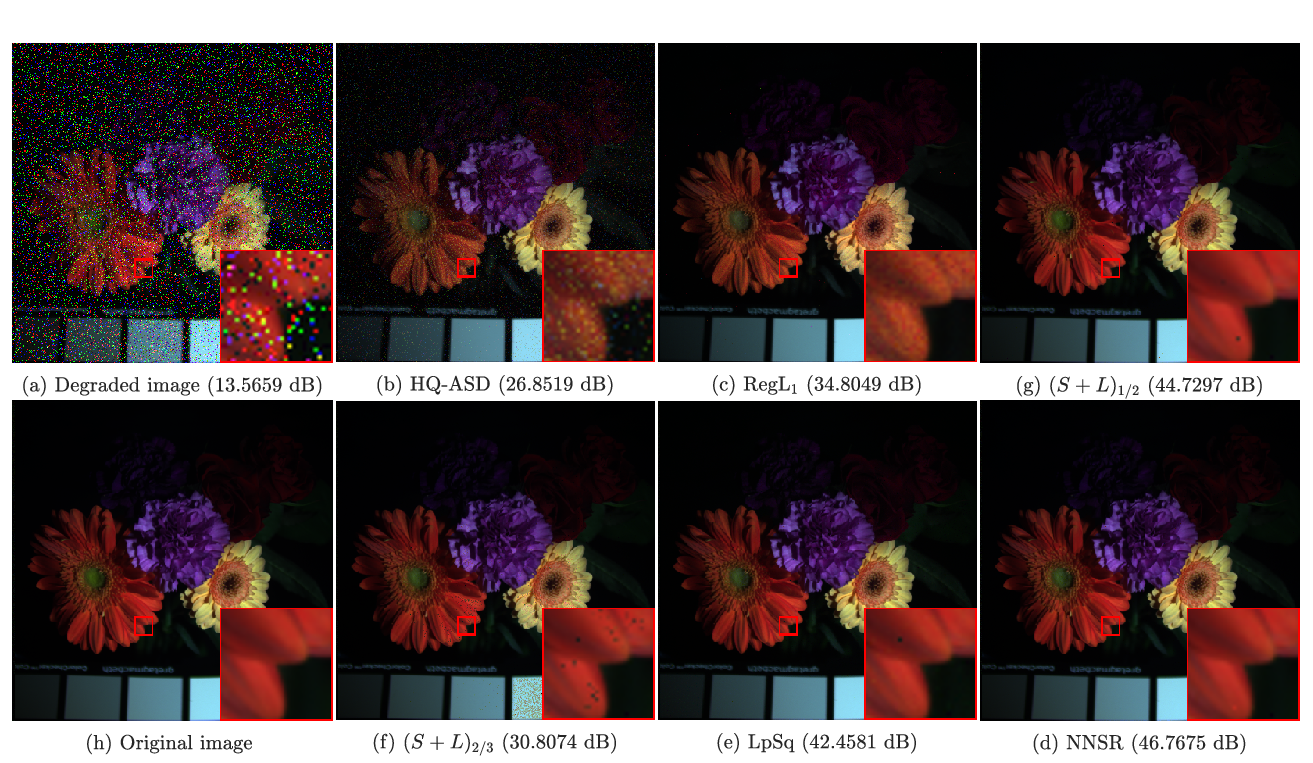}
		\vspace{-0.5em}
		\caption{Recovered images of `flowers' with bands 25-15-5 as R-G-B. (a) is the degraded image corrupted by impulsive noise and random mask, (h) is the original noise-free image, and the remaining images are the restoration results using different algorithms, with a demarcated area zoomed-in $6$ times.}\label{flowers_images}
	\end{figure} 
	
	Table~\ref{three_metrics} tabulates the recovery results in terms of PSNR, SSIM and runtime (in seconds). NNSR attains the highest PSNR and SSIM values for both datasets. $\rm RegL_1$ involves less running time than our method, but it requires knowing the matrix rank. On the other hand, LpSq and NNSR do not need the prior rank information, and compared with LpSq, NNSR has less computational time because LpSq involves iterations to find the proximal operator of the $\ell_p$-norm, while NNSR has a closed-form expression for its proximal operator. 
	
	Figs.~\ref{band_feather} and \ref{band_flower} show the recovered results for each band of `feathers' and `flowers', respectively. We observe that compared with the competing methods, NNSR has the highest PSNR and SSIM values for most of the bands in both datasets. Note that all techniques have a bad performance for the first few bands, because there exists a blur in these bands~\cite{DianR2018}. To provide visual comparison, three bands of MSI are chosen to construct a pseudo-color image.	Figs.~\ref{feather_images} and \ref{flowers_images} display the restoration results. It is seen that NNSR has the best recovery performance since the images generated by the remaining algorithms still contain apparent noise.

	\section{Conclusion}\label{Conclusion}
	In this paper, we devise a novel loss function, referred to as HOW, and propose a new sparsity-promoting regularizer, which is able to make the solution sparse. 
	Besides, the solution generated by our regularizer has less bias than that by the $\ell_1$-norm. Compared with the $\ell_p$ norm with $0<p<1$, the closed-form expression for the proximity operator of the developed regularizer is derived. Moreover, the properties of our regularizer are theoretically analyzed. We apply it to RMC, and an ADMM based algorithm with convergence guarantees is suggested. We prove that any generated accumulation point satisfies the KKT conditions.
	Extensive numerical examples using synthetic and real-world datasets show that our algorithm is superior to the state-of-the-art robust methods in terms of recovery performance. 

	\begin{appendices}		
		
		\section{~}\label{f-convex}
		\begin{IEEEproof}
			It is easy to find that $f(x)$ is a convex function if and only if $f(x)$ is convex when $x\textgreater \lambda$. Thus, we only need to verify that $f''(x)>0$ for $x \textgreater \lambda$. Then, we have:
			\begin{equation}\label{f-one-order}
				f^{'}(x) = x - x\cdot e^{\frac{\lambda^2-x^2}{\sigma^2}}
			\end{equation}
			and 
			\begin{equation}\label{f-two-order}
				\begin{split}
					f^{''}(x) &= 1-\left(1-\frac{2x^2}{\sigma^2}\right)e^{\frac{\lambda^2-x^2}{\sigma^2}} \\
					&= e^{\frac{\lambda^2-x^2}{\sigma^2}}\left(e^{\frac{x^2-\lambda^2}{\sigma^2}}+\frac{2x^2}{\sigma^2}-1\right)\\
					&\geq e^{\frac{\lambda^2-x^2}{\sigma^2}}\left(\frac{x^2-\lambda^2}{\sigma^2} +1 +\frac{2x^2}{\sigma^2}-1\right)\\
					&=e^{\frac{\lambda^2-x^2}{\sigma^2}}\left(\frac{3x^2-\lambda^2}{\sigma^2} \right)\\
					&> 0
				\end{split}
			\end{equation}
			where the first inequality is obtained because $e^x\textgreater x+1$ for any $x\in \mathbb{R}$, and the last inequality is due to $x\textgreater \lambda$. Thus, $f(x)$ is a convex function.
		\end{IEEEproof}

		{\section{Proof of Proposition~\ref{solution-proximal} \label{proof-solution-proximal}}
			\begin{IEEEproof}
				(i). When $y\textgreater 0$, the solution to ${\rm arg}\mathop{\max}\limits_{x} ~y\cdot x -f(x)$ is unique, denoted as $x^\star$, and it satisfies via (\ref{y-pro-solution}):
				\begin{equation}\label{slope-y}
					y = x^\star - x^\star e^{\frac{\lambda^2-(x^\star)^2}{\sigma^2}}
				\end{equation}
				implying that $y$ increases with $x^\star$ because the right hand side is monotonically increasing w.r.t. $x^\star$ via (\ref{f-one-order}) and (\ref{f-two-order}). 
				Besides, using Lemma~\ref{solution-proximal} yields:
				\begin{equation}\label{key-ralation}
					{\rm arg}\mathop{\max}\limits_{x} ~xy -f(x) = \partial f^*(y)=y + \lambda \partial \varphi_{\sigma,\lambda}(y)
				\end{equation}
				that is, $x^\star = y + \lambda\partial \varphi_{\sigma,\lambda}(y)$. By (\ref{slope-y}), it is easy to get:
				\begin{equation}
					\partial \varphi_{\sigma,\lambda}(y) = \frac{x^\star}{\lambda}e^{\frac{\lambda^2-(x^\star)^2}{\sigma^2}}
				\end{equation}
				We define $r(x)= \frac{x}{\lambda} e^{\frac{\lambda^2-x^2}{\sigma^2}}$ and $r'(x) = \frac{1}{\lambda}\left(1-\frac{2x^2}{\sigma^2}\right)\times e^{\frac{\lambda^2-x^2}{\sigma^2}}<0$ because $\frac{2x^2}{\sigma^2}>1$ when $x>\lambda$, namely, $\sigma\leq \sqrt{2}\lambda$. Combining (\ref{slope-y}), that is, $y$ increases with $x^\star$, we know that $\partial \varphi_{\sigma,\lambda}(y)$ is a decreasing function of $y$. Thus, $\varphi_{\sigma,\lambda}(y)$ is concave for $y>0$.
				
				Besides, according to (\ref{varphi-y}), we have:
				\begin{equation}\label{symmetric-property}
					\begin{split}
						\varphi_{\sigma,\lambda}(-y) &= \mathop {\max}\limits_{x\in \mathbb{R}}~ -\frac{(-y-x)^2}{2\lambda} + \frac{l_{\sigma,\lambda}(x)}{\lambda}\\
						&\overset{t=-x}{=}\mathop {\max}\limits_{t\in \mathbb{R}}~ -\frac{(-y+t)^2}{2\lambda} + \frac{l_{c,\sigma}(-t)}{\lambda}\\
						&=\mathop {\max}\limits_{t\in \mathbb{R}}~ -\frac{(y-t)^2}{2\lambda} + \frac{l_{c,\sigma}(t)}{\lambda}\\
						& = \varphi_{\sigma,\lambda}(y)
					\end{split}
				\end{equation}
				where the penultimate equation is obtained because $l_{\sigma,\lambda}(x)$ is an even function. Therefore, $\varphi(y)$ is symmetric.
				
				(ii).  Due to the fact that the conjugate of $f(x)$ is a convex function, we know that $f^*(y) = \lambda \varphi_{\sigma,\lambda}(y) +\frac{y^2}{2}$ in (\ref{fy-LF}) is convex w.r.t. $y$. Thus, $g(y)$ is convex w.r.t. $y$ for any $\lambda$ and $\sigma$.
				
				(iii). Since $P_{\varphi_{\sigma,\lambda}}(x)$ is an odd function, and $P_{\varphi_{\sigma,\lambda}}(x) =0$ when $|x|\leq \lambda$, we only need to verify that $P_{\varphi_{\sigma,\lambda}}(x) = x - x\cdot e^{(\lambda^2-x^2)/\sigma^2}$ is monotonic when $x>\lambda$. It is easy to conclude that $P_{\varphi_{\sigma,\lambda}}(x)$ is monotonically increasing when $x>\lambda$ via (\ref{f-one-order}) as well as (\ref{f-two-order}), thus $P_{\varphi_{\sigma,\lambda}}(x)$ is monotonically non-decreasing.

		\end{IEEEproof}} 
		
		\section{Proof of Theorem~\ref{theorem-convergence} \label{proof-theorem-convergence}}
		The following two propositions are first provided.
		\begin{mypro}\label{bias-comparison} 
			When $\sigma \leq \sqrt{2}\lambda$, namely, $\varphi_{\sigma,\lambda}(y)$ is concave for $y>0$, $|P_{\ell_1,\lambda}(x)| \leq |P_{\varphi_{\sigma,\lambda}}(x)|$ and $|x-P_{\varphi_{\sigma,\lambda}}(x)| \leq \lambda$ for any $x\in \mathbb{R}$, implying that the bias generated by our regularizer is no more than that by the $\ell_1$-norm.
		\end{mypro}
		\begin{IEEEproof}
			Both $P_{\ell_1,\lambda}(x)$ and $P_{\varphi_{\sigma,\lambda}}(x)$ are odd functions, and according to (\ref{Pro-L1}) and (\ref{y-pro-solution}), we only need to verify $P_{\ell_1,\lambda}(x) \leq P_{\varphi_{\sigma,\lambda}}(x)$ when $x\geq \lambda$. Thus, when $x\geq \lambda$, we have:
			\begin{equation}
				\begin{split}
					\Delta(x) &= P_{\varphi_{\sigma,\lambda}}(x)- P_{\ell_1,\lambda}(x)=-x\cdot e^{(\lambda^2-x^2)/\sigma^2} + \lambda
				\end{split}
			\end{equation}
			It is easy to check that $\Delta(x)$ increases with $x$ when $\sigma \leq \sqrt{2}\lambda$, and $\Delta(x)\geq \Delta(\lambda)=0$. Thus, $P_{\ell_1,\lambda}(x) \leq P_{\varphi_{\sigma,\lambda}}(x)$, $x-P_{\varphi_{\sigma,\lambda}}(x) \leq x-P_{\ell_1,\lambda}(x)= \lambda$ for $x\geq \lambda$ and $x-P_{\varphi_{\sigma,\lambda}}(x) = x-P_{\ell_1,\lambda}(x)\leq \lambda$ for $0<x<\lambda$. Therefore, $|x-P_{\varphi_{\sigma,\lambda}}(x)| \leq \lambda$ for any $x\in \mathbb{R}$.	
		\end{IEEEproof}
		\begin{mypro}\label{phi_lambda} 
			Defining $h(\sigma,\lambda)=\lambda \varphi_{\sigma,\lambda}(y)$, then when $y>0$, $h(\sigma,\lambda)$ increases with $\lambda$ and $\sigma$.
		\end{mypro}
		\begin{IEEEproof}
			According to (\ref{varphi-y}), we have $h(\sigma,\lambda) = -\frac{(y-x^\star)^2}{2} + l_{\sigma,\lambda}(x^\star)$. By (\ref{y-pro-solution}), we know $x>\lambda$ for $y>0$, thus we only need to verify that $h(\sigma,\lambda)$ increases with $\lambda$ and $\sigma$ when $x^\star>\lambda$. We can check that $\frac{\partial h}{\partial \lambda}=\lambda(1-e^{\frac{\lambda^2-(x^\star)^2}{\sigma^2}})>0$ and $\frac{\partial h}{\partial \sigma}=(\sigma e^{\frac{(x^\star)^2-\lambda^2}{\sigma^2}} -\sigma +{\frac{\lambda^2-(x^\star)^2}{\sigma}} )e^{\frac{\lambda^2-(x^\star)^2}{\sigma^2}}> (\sigma ({\frac{(x^\star)^2-\lambda^2}{\sigma^2}}+1) -\sigma +{\frac{\lambda^2-(x^\star)^2}{\sigma}} )e^{\frac{\lambda^2-(x^\star)^2}{\sigma^2}}=0$. This completes the proof.	
		\end{IEEEproof}
		\subsection*{Proof of Theorem~\ref{theorem-convergence}:}
		\begin{IEEEproof}
			(i). We first prove the boundness of $\pmb \Lambda^{k+1}$ via:
			\begin{equation}\label{bounded_Lambda}
				\begin{split}
					\left\|\pmb \Lambda^{k+1}\right\|_F^2&= \left\|\pmb \Lambda^{k}+\rho^k\left(\pmb X -\pmb M^{k+1}-\pmb S^{k+1}\right)\right\|_F^2\\
					&={\left(\rho^k\right)^2}\left\|\pmb X -\pmb M^{k+1}+\frac{\pmb \Lambda^{k}}{\rho^k} -\pmb S^{k+1}\right\|_F^2\\
					&\overset{a}{=}{\left(\rho^k\right)^2}\left\|\pmb X_\Omega -\pmb M_\Omega^{k+1}+\frac{\pmb \Lambda_\Omega^{k}}{\rho^k} -\pmb S_\Omega^{k+1}\right\|_F^2\\
					&{=}{\left(\rho^k\right)^2}\left\|\pmb D_\Omega^{k+1}  - P_{\varphi_{\cdot,\lambda/\rho^k}}\left(\pmb D_\Omega^{k+1}\right)\right\|_F^2\\
					&\overset{b}{\leq} {\left(\rho^k\right)^2} \sum_{i=1}^{|\Omega|_1} (\lambda/\rho^k)^2\\
					&=|\Omega|_1 \lambda^2\\
				\end{split}
			\end{equation}
			where $\pmb D_\Omega^{k+1} = \pmb X_\Omega -\pmb M_\Omega^{k+1}+\frac{\pmb \Lambda_\Omega^{k}}{\rho^k}$, $a$ and $b$ are owing to (\ref{S_solution_Om_S}) and Proposition~\ref{bias-comparison}, respectively. Thus, $\left\|\pmb \Lambda^{k+1}\right\|_F$ is bounded from above.
			
			Besides, by (\ref{M_solution}) and (\ref{ADMM_Ladm}), we obtain:
			\begin{equation}\label{Fro_con_M}
				\begin{split}
					&\lim_{k\rightarrow \infty}\left\|\pmb M^{k+1}-\pmb M^{k}\right\|_F^2= \lim_{k\rightarrow \infty}\|P_{\|\cdot\|_{\varphi_{\sigma,1/\rho^k}}}\left(\pmb X-\pmb S^{k} + {\pmb \Lambda^k}/{\rho^k}\right) \\ &~~~~~~~~~~~~~~~~~~~~~~~~	-\left(\pmb X - \pmb S^k-{(\pmb \Lambda^k - \pmb \Lambda^{k-1})}/{\rho^{k-1}}\right)\|_F^2\\
					&= \lim_{k\rightarrow \infty}\|P_{\|\cdot\|_{\varphi_{\sigma,1/\rho^k}}}(\pmb Y^k ) -\pmb Y^k +\pmb \Lambda^k /{\rho^k}+{(\pmb \Lambda^k - \pmb \Lambda^{k-1})}/{\rho^{k-1}}\|_F^2\\
					&\leq \lim_{k\rightarrow \infty}\|P_{\|\cdot\|_{\varphi_{\sigma,1/\rho^k}}}(\pmb Y^k ) -\pmb Y^k \|_F^2 +\lim_{k\rightarrow \infty}\|\pmb Z^k\|_F^2\\
					&= \lim_{k\rightarrow \infty}\|P_{{\varphi_{\sigma,1/\rho^k}}}(\pmb s) - \pmb s\|_F^2 +\lim_{k\rightarrow \infty}\|\pmb Z^k\|_F^2\\
					&\overset{c}{\leq} \lim_{k\rightarrow \infty}|\pmb s|_0/\rho^k +\lim_{k\rightarrow \infty}\|\pmb Z^k\|_F^2 =0
				\end{split}
			\end{equation}
			where $\pmb Y^k = \pmb X-\pmb S^{k} + {\pmb \Lambda^k}/{\rho^k}=\pmb U~{\rm diag}(P_{{\varphi_{\sigma,1/\rho^k}}}(\pmb s))~\pmb V^T$ with $\pmb s$ being the singular value vector of $\pmb Y^k$, $|\pmb s|_0$ represents the number of nonzero elements of $\pmb s$, $\pmb Z^k = \pmb \Lambda^k /{\rho^k}+{(\pmb \Lambda^k - \pmb \Lambda^{k-1})}/{\rho^{k-1}}$ and (c) is owing to Proposition~\ref{bias-comparison}.
			
			Similar to (\ref{Fro_con_M}), we have:
			\begin{equation}
				\lim_{k\rightarrow \infty}\left\|\pmb S^{k+1}-\pmb S^{k}\right\|_F^2=0
			\end{equation}
			Moreover, by (\ref{ADMM_Ladm}), we get:
			\begin{equation}\label{Proof_|ambda_k}
				\lim_{k\rightarrow \infty}\left\| \pmb X -\pmb M^{k+1}-\pmb S^{k+1} \right\|_F^2 = \lim_{k\rightarrow \infty}\| \pmb \Lambda^{k+1}-\pmb \Lambda^{k} \|_F^2/{\rho^k}=0
			\end{equation}
			which means that the generated sequence $\{(\pmb M^k, \pmb S^k)\}$ is a feasible solution to the objective function.
			
			(ii). Since $\pmb M^k$ and $\pmb S^k$ are the minimizers of their corresponding optimization problems, we have the following inequalities:
			\begin{equation}\label{Proof_update_E}
				\begin{split}
					\mathcal{L}_{\rho^k}(\pmb M^{k+1}, \pmb S^{k},\pmb \Lambda^k)&\leq \mathcal{L}_{\rho^k}(\pmb M^k, \pmb S^k,\pmb \Lambda^k)\\
					\mathcal{L}_{\rho^k}(\pmb M^{k+1}, \pmb S^{k+1},\pmb \Lambda^k)&\leq \mathcal{L}_{\rho^k}(\pmb M^{k+1}, \pmb S^k,\pmb \Lambda^k)
				\end{split}
			\end{equation}
			Besides, we have:
			\begin{equation*}
				\begin{split}
					&\mathcal{L}_{\rho^{k+1}}(\pmb M^{k+1}, \pmb S^{k+1},\pmb \Lambda^{k+1})\overset{d}{\leq} \mathcal{L}_{\rho^k}(\pmb M^{k+1}, \pmb S^{k+1},\pmb \Lambda^k)\\
					&+\left<\frac{\pmb \Lambda^{k+1}}{\rho^{k+1}} - \frac{\pmb \Lambda^{k}}{\rho^{k}},\pmb X -\pmb M^{k+1}-\pmb S^{k+1} \right>\\
					&=\mathcal{L}_{\rho^k}(\pmb M^{k+1}, \pmb S^{k+1},\pmb \Lambda^k)+ \left<\frac{\pmb \Lambda^{k+1}}{\rho^{k+1}} - \frac{\pmb \Lambda^{k}}{\rho^{k}},\frac{\pmb \Lambda^{k+1}-\pmb \Lambda^{k}}{\rho^k} \right>
				\end{split}
			\end{equation*}
			where $d$ is due to Proposition~\ref{phi_lambda}, and
			\begin{equation*}
				\begin{split}
					&\left<\frac{\pmb \Lambda^{k+1}}{\rho^{k+1}} - \frac{\pmb \Lambda^{k}}{\rho^{k}},\frac{\pmb \Lambda^{k+1}-\pmb \Lambda^{k}}{\rho^k} \right>\\
					=&1/(\rho^k)^2 \left<\pmb \Lambda^{k+1}/\mu- \pmb \Lambda^{k},\pmb \Lambda^{k+1}-\pmb \Lambda^{k} \right>\\
					=&1/(\rho^k)^2 \left(\|\pmb \Lambda^{k+1}\|_F^2/\mu + \|\pmb \Lambda^{k}\|_F^2 - (1+1/\mu)\left<\pmb \Lambda^{k+1} - \pmb \Lambda^{k}\right>\right)\\
					\leq&1/(\rho^k)^2 \left(\|\pmb \Lambda^{k+1}\|_F^2/\mu + \|\pmb \Lambda^{k}\|_F^2 \right. \\ & \left. + (1+1/\mu)/2(\|\pmb \Lambda^{k+1}\|_F^2+\|\pmb \Lambda^{k}\|_F^2)\right)\\
					\leq& 1/(\rho^k)^2(2+2/\mu)|\Omega|_1 \lambda^2
				\end{split}
			\end{equation*}
			Hence, 
			\begin{equation}\label{Proof_update_rho}
				\begin{split}
					\mathcal{L}_{\rho^{k+1}}(\pmb M^{k+1}, \pmb S^{k+1},\pmb \Lambda^{k+1})&\leq \mathcal{L}_{\rho^k}(\pmb M^{k+1}, \pmb S^{k+1},\pmb \Lambda^k)\\
					&+1/(\rho^k)^2(2+2/\mu)|\Omega|_1 \lambda^2
				\end{split}
			\end{equation}
			Combining (\ref{Proof_update_E}) and (\ref{Proof_update_rho}) yields:
			\begin{equation}\label{Proof_update_all}
				\begin{split}
					\mathcal{L}_{\rho^{k+1}}(\pmb M^{k+1}, \pmb S^{k+1},\pmb \Lambda^{k+1})&\leq \mathcal{L}_{\rho^k}(\pmb M^k, \pmb S^k,\pmb \Lambda^k)\\
					&+1/(\rho^k)^2(2+2/\mu)|\Omega|_1 \lambda^2
				\end{split}
			\end{equation}
			Thus, we get:
			\begin{equation}\label{Proof_update_acc}
				\begin{split}
					\mathcal{L}_{\rho^{k}}(\pmb M^{k}, \pmb S^{k},\pmb \Lambda^{k})\leq &\mathcal{L}_{\rho^0}(\pmb M^0, \pmb S^0,\pmb \Lambda^0)\\
					&+(2+2/\mu)|\Omega|_1 \lambda^2\sum_{i=0}^{k-1}1/(\rho^i)^2
				\end{split}
			\end{equation}
			Given a bounded initialization, since $\lim_{k\rightarrow \infty}\sum_{i=0}^{k-1}1/(\rho^i)^2 = \frac{\mu^2}{(\rho^0)^2(\mu^2-1)}<\infty$, we conclude that $\mathcal{L}_{\rho^{k}}(\pmb M^{k}, \pmb S^{k},\pmb \Lambda^{k})$ is bounded from above. We then know that $\mathcal{L}_{\rho^k}(\pmb M^k, \pmb S^{k+1},\pmb \Lambda^k)$ and $\mathcal{L}_{\rho^k}(\pmb M^{k+1}, \pmb S^{k+1},\pmb \Lambda^k)$ are bounded from above via (\ref{Proof_update_E}), implying that the sequences $\{(\pmb S^{k+1}, \pmb M^{k+1})\}$ are bounded. This is because if $\|\pmb S^{k+1}\|_F^2\rightarrow \infty$ or $\|\pmb M^{k+1}\|_F^2\rightarrow \infty$ at the $(k+1)$th iteration, then $\mathcal{L}_{\rho^k}(\pmb M^k, \pmb S^{k+1},\pmb \Lambda^k)\rightarrow \infty$ or $\mathcal{L}_{\rho^k}(\pmb M^{k+1}, \pmb S^{k+1},\pmb \Lambda^k)\rightarrow \infty$. 
			
			Therefore, combining (\ref{bounded_Lambda}), we conclude that the sequences $\{(\pmb M^k, \pmb S^k, \pmb \Lambda^k)\}$ are all bounded.
			
			(iii). By Bolzano-Weierstrass theorem~\cite{BartleRG2011}, the boundedness of $\{(\pmb M^k, \pmb S^k, \pmb \Lambda^k)\}$ guarantees that there exists at least one accumulation point $({\pmb M^*}, {\pmb S^*}, {\pmb \Lambda^*})$ for $\{(\pmb M^k, \pmb S^k, \pmb \Lambda^k)\}$. That is, there exists a convergent subsequence $\{(\pmb M^{k_j}, \pmb S^{k_j}, \pmb \Lambda^{k_j})\}$ such that
			\begin{subequations}
				\begin{align}
					\lim_{k_j\rightarrow \infty} \pmb S^{k_j} &= {\pmb S^*}\label{EkJ}\\
					\lim_{k_j\rightarrow \infty} \pmb M^{k_j} &= {\pmb M^*}\label{MKJ}\\
					\lim_{k_j\rightarrow \infty} \pmb \Lambda^{k_j} &= {\pmb \Lambda^*}\label{LKJ}
				\end{align}
			\end{subequations} 
			
			In addition, the KKT conditions for (\ref{RMC-formulation}) are:
			\begin{subequations}
				\begin{align}
					&\pmb X = \pmb M^* + \pmb S^*\label{KKTX}\\
					& \pmb \Lambda^* \in \partial \|\pmb M^*\|_{\varphi_{\sigma,1/\rho^*}} \label{KKTM}\\
					&\pmb \Lambda_\Omega^* \in \lambda \partial \varphi_{\sigma,\lambda/\rho^*}(\pmb S^*_\Omega)\label{KKTS}
				\end{align}
			\end{subequations} 
			As $\{\pmb \Lambda^k\}$ is bounded, (\ref{KKTX}) is satisfied due to:
			\begin{equation}
				\begin{split}
					\left\|\pmb X - \pmb M^* - \pmb S^*\right\|_F^2 &= \lim_{k_j\rightarrow \infty} \left\|\pmb X - \pmb M^{k_j+1} - \pmb S^{k_j+1}\right\|_F^2\\
					&=\lim_{k_j\rightarrow \infty} \left\| \pmb \Lambda^{k_j+1}-\pmb \Lambda^{k_j} \right\|_F^2/\rho^{k_j}\\
					& = 0
				\end{split}
			\end{equation}
			Besides, $\pmb M^{k+1}$ and $\pmb S^{k+1}$ calculated by (\ref{M_solution}) and (\ref{S_solution}) are the minimizers for their corresponding optimization problems, thus we have:
			\begin{subequations}
				\begin{align}
					\pmb 0 \in \frac{\partial \mathcal{L}(\pmb M^{k+1}, \pmb S^{k},\pmb \Lambda^k)}{\partial \pmb M}\label{parM}\\
					\pmb 0 \in \frac{\partial \mathcal{L}(\pmb M^{k+1}, \pmb S^{k+1},\pmb \Lambda^k)}{\partial \pmb S}\label{prtS}
				\end{align}
			\end{subequations}
			Moreover, 
			\begin{equation}
				\begin{split}
					\pmb 0 &\in \frac{\partial \mathcal{L}(\pmb M^{k+1}, \pmb S^{k},\pmb \Lambda^k)}{\partial \pmb M}\\
					& = \partial \|\pmb M^{k+1}\|_{\varphi_{\sigma,1/\rho^k}} - \pmb \Lambda^k - \rho^k (\pmb X - \pmb M^{k+1}-\pmb S^k)\\
					& = \partial \|\pmb M^{k+1}\|_{\varphi_{\sigma,1/\rho^k}} - \pmb \Lambda^{k+1} - \rho^k (\pmb S^{k+1}-\pmb S^k)
				\end{split}
			\end{equation}
			Hence, we have:
			\begin{equation}\label{proof_M}
				\begin{split}
					\pmb 0 &\in  \lim_{k_j\rightarrow \infty} \partial \|\pmb M^{k_j+1}\|_{\varphi_{\sigma,{1/{\rho^{k_j}}}}} - \pmb \Lambda^{k_j+1} - \rho^{k_j} (\pmb S^{k_j+1}-\pmb S^{k_j})\\
					& = \partial \|\pmb M^*\|_{\varphi_{\sigma,1/\rho^*}} - \pmb \Lambda^*
				\end{split}
			\end{equation}
			and (\ref{KKTM}) is proved.
			Furthermore,
			\begin{equation}
				\begin{split}
					\pmb 0 &\in \frac{\partial \mathcal{L}(\pmb M^{k+1}, \pmb S^{k+1},\pmb \Lambda^k)}{\partial \pmb S_\Omega}\\
					& = \lambda \partial \varphi_{\sigma,\lambda/\rho^k}(\pmb S^{k+1}_\Omega) - \pmb \Lambda_\Omega^k - \rho^k (\pmb X_\Omega - \pmb M_\Omega^{k+1}-\pmb S_\Omega^{k+1})\\
					& = \partial \|\pmb M^{k+1}\|_{\varphi_{\sigma,1/\rho^k}} - \pmb \Lambda_\Omega^{k+1}
				\end{split}
			\end{equation}
			thus, we get:
			\begin{equation}\label{proof_SO}
				\begin{split}
					\pmb 0 &\in \lim_{k_j\rightarrow \infty}  \partial \|\pmb M^{k_j+1}\|_{\varphi_{\sigma,1/\rho^{k_j}}} - \pmb \Lambda_\Omega^{k_j+1}\\
					& = \partial \|\pmb M^{*}\|_{\varphi_{\sigma,1/\rho^*}} - \pmb \Lambda_\Omega^{*}
				\end{split}
			\end{equation}
			and (\ref{KKTS}) is satisfied.
			
			Therefore, any accumulation point $\{{\pmb M^*}, {\pmb S^*}, {\pmb \Lambda^*}\}$ satisfies the KKT conditions and is a stationary point. 
			
		\end{IEEEproof}

	\end{appendices}

\end{document}